\documentclass{aastex62}

\graphicspath{{./}{figures/}}

\received{August xx, 2019}
\revised{\today}
\accepted{October zz, 2019}
\submitjournal{ApJ}

\shorttitle{Influence of density dependence}
\shortauthors{Sumiyoshi et al.}

\begin{document}

\title{Influence of density dependence of symmetry energy in hot and dense matter for supernova simulations}

\correspondingauthor{Kohsuke Sumiyoshi}
\email{sumi@numazu-ct.ac.jp}

\author[0000-0002-9224-9449]{Kohsuke Sumiyoshi}
\affiliation{National Institute of Technology, Numazu College, Shizuoka 410-8501, Japan}

\author[0000-0001-6330-1685]{Ken'ichiro Nakazato}
\affiliation{Faculty of Arts and Science, Kyushu University, Fukuoka 819-0395, Japan}

\author{Hideyuki Suzuki}
\affiliation{Faculty of Science and Technology, Tokyo University of Science, Noda, Chiba 278-8510, Japan}

\author{Jinniu Hu}
\affiliation{School of Physics, Nankai University, Tianjin 300071,  China}

\author{Hong Shen}
\affiliation{School of Physics, Nankai University, Tianjin 300071,  China}



\begin{abstract}
We study the influence of density-dependent symmetry energy at high densities in simulations of core-collapse supernovae, black hole formation and proto-neutron star cooling by extending the relativistic mean field (RMF) theory used for the Shen EOS table.  
We adopt the extended RMF theory to examine the density dependence of the symmetry energy with a small value of the slope parameter $L$, while the original properties of the symmetric nuclear matter are unchanged.  
In order to assess matter effects at high densities, we perform numerical simulations of gravitational collapse of massive stars adopting the EOS table at high densities beyond $10^{14}$ g/cm$^3$ with the small $L$ value, which is in accord with the experimental and observational constraints,  and compare them with the results obtained by using the Shen EOS.  
Numerical results for 11.2M$_{\odot}$ and 15M$_{\odot}$ stars exhibit minor effects around the core bounce and in the following evolution for 200 ms.  
Numerical results for 40M$_{\odot}$ and 50M$_{\odot}$ stars reveal a shorter duration toward the black hole formation with a smaller maximum mass for the small $L$ case.  
Numerical simulations of proto-neutron star cooling over 10 s through neutrino emissions demonstrate increasing effects of the symmetry energy at high densities.  
Neutrino cooling drastically proceeds in a relatively long timescale with high luminosities and average energies with the small symmetry energy.  
Evolution toward the cold neutron star is affected because of the different behavior of neutron-rich matter while supernova dynamics around core bounce remains similar in less neutron-rich environments.  
\end{abstract}



\section{Introduction} \label{sec:intro}

Properties of hot and dense matter are central to every stage of the stellar death and the birth of compact objects.  
The fate of massive stars relies on the equation of state (EOS) of highly compressed matter, which determines explosion or collapse through stellar stability and hydrodynamics.  
Core-collapse supernovae are driven by sudden repulsion of matter beyond the nuclear matter density in the gravitational collapse of iron core at the center of massive stars \citep{bet90}.  
Outcome of the propagating shock wave, which is launched by the bounce of central core, depends on the thermodynamical properties of hot and dense matter in a non-trivial way of neutrino-matter coupled problems \citep{kot12,jan12a,bur13,jan17b}.  
The birth of compact objects (neutron star or black hole) is crucially determined by the EOS through its maximum stable configuration as the threshold.  Hence, the role of dense matter in core-collapse supernovae has been one of remaining issues in the study of explosion mechanism for many decades \citep{oer17}.  

The role of EOS in supernovae has been explored along with the progress of studies on hot and dense matter because of many aspects of uncertainties at the wide variety of extreme conditions.  
Early studies focused on the behavior of uniform matter around the nuclear matter density in terms of parameters in a functional form of EOS.  
They systematically clarified the role of stiffness of EOS on explosions through the degree of compression during the core bounce \citep{bar85,tak88}.  
It has become an intuitive view that a softer EOS leads to a higher compression to release a larger gravitational energy, which is favorable for successful explosions.  

It has been a difficult task to investigate the influence in greater detail due to the limitation of EOS designed for numerical simulations for hydrodynamics and neutrino transport.  
Main complication is consistent description of uniform and non-uniform matter to provide various quantities of thermodynamics by covering the wide range of density, composition and temperature.  
One has to construct the framework to describe nuclear matter from nuclear interactions and to apply it to the various phases of matter as gas and liquid.  
Such construction of the EOS for supernovae has been made by extension of \citet{bar85,bet79} based on the non-relativistic effective nuclear interaction \citep{lat91} (See also \citep{hil84}) and has been uniquely used in supernova simulations at the early stage.  
The influence of the EOS on the early phase of evolution of shock wave after the bounce has been studied by utilizing the choice of the incompressibility \citep{tho03} (See also \citet{bru89a,swe94} for symmetry energy).  

Additional construction of the EOS table for supernova simulations has been later made in the relativistic mean field (RMF) theory \citep{she98a,she98b} in addition to the Lattimer-Swesty EOS.  
The Shen EOS has been used to examine the influence of the EOS \citep{sum05} having a  different form of supernova EOS.  
It has been demonstrated that the EOS effect at high densities is not so drastic and evolves in a delicate manner through the coupling of  neutrinos and matter.  
Properties of non-uniform matter at low densities are also important since mixed composition of nucleons and nuclei is essential to delineate the reactions and transport of neutrinos.  
The neutrino heating mechanism \citep{bet85}, which is most credible among various scenarios of explosions, proceeds through emissions of neutrinos from the compact object and absorption of neutrinos in ambient material behind the shock wave.  

Extensive studies of the supernova EOS have been made by extending these frameworks to investigate the EOS effects in the neutrino-driven supernovae.  
Revisions of the RMF theory have been made by modifications of the effective nuclear interaction to improve the agreement with the nuclear experiments and neutron star observations \citep{gshe11,hem12,ste13}.  
The mixture of nuclei instead of a simple approximation with single species of nuclei has been described in modern sets of supernova EOS tables \citep{hem10,fur11,fur17a}.  
Microscopic approaches of nuclear many-body theories have been also applied to construct the EOS data table \citep{tog17,fur17b,sch19a,fur19}. 
Exploration in the variations of EOS is also made by changing parameters in the functional form of the Lattimer-Swesty EOS \citep{sch17,sch19b,yas18}.  
(See \citep{oer17} for a comprehensive list of the supernova EOS)

These sets of supernova EOS have been utilized to explore the influence of EOS in the numerical simulations of core-collapse supernovae.  
Under the spherical symmetry, the EOS influence on dynamics from core bounce to stall of shock wave \citep{sum05,hem12,ste13} and the formation of black holes \citep{sum06,sum07,nak10a,ocon11} from massive stars has been studied in detail.  
In these studies, the emission of neutrinos is also examined in the variations of the EOS.  
Supernova neutrinos from the cooling of proto-neutron stars have been studied to examine the EOS effect by adopting the EOS tables \citep{sum95c,pon99,nak18,nak19}.  
The effect of symmetry energy on convection in proto-neutron stars is studied and its associated feature in neutrino emission is proposed \citep{rob12}.  
Since the importance of hydrodynamical instabilities has been elucidated in the explosion mechanism, the role of EOS has been studied in numerical simulations in multi-dimensions \citep{mar09b,jan12a,suw13,cou13,fis14,pan18,nag18,sch19b}.  
In these studies in 2D, common tendency of favorable conditions for explosion is implied for a softer EOS in the available sets.  
A systematic modification of the effective mass has been recently explored in 3D simulations \citep{sch19b}.  
Further studies on EOS influence are required to sort out intertwined processes of the neutrino transport and hydrodynamical instabilities.   

We focus on the extension of EOS tables by \citet{she98a,she98b,she11} with modifications of nuclear interactions and explore influence of the density-dependent symmetry energy on the core-collapse supernovae and proto-neutron stars.  
The Shen EOS tables have been widely adopted in astrophysical simulations together with the Lattimer-Swesty EOS as one of the benchmark tests.  
It is also used to provide a set of the prediction of neutrino burst in the supernova neutrino database by numerical simulations of core-collapse supernovae, proto-neutron star cooling and the black hole formation \citep{sum07,nak13a}.  
The RMF theory used in the Shen EOS is extended to develop the EOS tables by including extra degrees of freedom of pions, hyperons and quarks \citep{ish08,nak08a,nak10b,nak12}.  
The extended version of the EOS table with the  mixture of nuclei is constructed by describing the multi-composition under the nuclear statistical equilibrium \citep{fur11,fur13a,fur17a} and is applied to the multi-dimensional core-collapse simulations \citep{nag18}.  

While there are wide applications to simulations, one of the issues to improve the series of Shen EOS tables is strong influence of the symmetry energy, which is common in the relativistic many-body theories.  
It has been often pointed out that the value of the symmetry energy of Shen EOS at the nuclear saturation is large as compared with the nuclear experiments.  
Furthermore the resulting neutron star EOS provides too large radii as compared with extracted values of neutron star radii from X-ray observations (see \citep{for15} for example).  
The detection of gravitational wave from neutron star merger GW170817 provides new constraints on the tidal deformability \citep{abb18}, which also restricts the neutron star radii, statistically excluding the predictions of neutron stars by Shen EOS.  
Although isovector part of the nuclear interactions of RMF extended from the original Shen EOS has been studied \citep{she97} despite the lack of experimental data at that time, there is room for improvement of the density dependence of symmetry energy with the current acquaintance.  

Recently, the density dependence of the symmetry energy in the RMF has been studied systematically \citep{bao14a,bao14b,bao15,hu19}.  
They extended the RMF by adding a coupling term between vector meson and isovector-vector meson to assess the effects of the symmetry energy while the properties of symmetric matter are unchanged.  
This modification is advantageous to explore the properties of neutron-rich matter for neutron stars and supernovae since the properties of finite nuclei are well reproduced by the RMF with the TM1 parameterization for the Shen EOS.  
They explored effects of the symmetry energy on properties of neutron star crusts 
by systematically changing the slope parameter of symmetry energy, $L$, from 110.8 MeV to 40 MeV.  
By changing the value of the slope parameter with the fixed symmetry energy at a certain density, it is possible to explore the density dependence of the symmetry energy at low and high densities with a satisfactory description of nuclei.  

We utilize this method of modification to examine the behavior of symmetry energy used in the Shen EOS.  
It was pointed out that the slope parameter of the RMF with TM1, $L=110.8$ MeV, is too large as compared to the recent nuclear experiments of symmetry energy \citep{bal16,li19} (See also \citet{tami14} on experiments, for example).  
By choosing the low value of slope parameter, $L=40$ MeV, which is well within the experimental constraints, one can provide a slow increase of the symmetry energy in contrast to the rapid increase in the original Shen EOS under the same behavior of symmetric nuclear matter.  
It also results in a small radius of neutron stars as well as a suitable value of the tidal deformability within the observational constraints (See also \citep{hu19} for details) while the maximum mass remains similar.  

In order to explore the density dependence of the symmetry energy in astrophysical simulations, we construct a data table of EOS for uniform matter at finite temperature with the same modification to $L$ value from the RMF with TM1.  
We prepare a modified Shen EOS table with low $L$ value for uniform matter at densities above $\sim10^{14}$ g/cm$^{3}$ with the usage of the Shen EOS for low densities, where non-uniform matter of nucleons appears.  
By switching only the uniform matter, we clarify the influence of the symmetry energy at high densities with respect to the results adopting the original Shen EOS.  
Revisions of the Shen EOS table for the whole density range with the Thomas-Fermi calculations and the nuclear statistical equilibrium are undertaken by Shen et al. and Furusawa et al. and will be published elsewhere.  
Influence of the symmetry energy from uniform and non-uniform matter will be studied by adopting the revised Shen EOS table in future.  

By adopting the two EOS tables, the modified Shen EOS with $L=40$ MeV and the original Shen EOS, we study the influence of the density dependence of symmetry energy on core-collapse supernovae, black hole formation and proto-neutron stars.  
By switching uniform matter with low and high $L$ values, we solely extract the effects of symmetry energy in central core under the same condition of outer layer of non-uniform matter where the information of mixture of nucleons and nuclei is also important.  
We perform spherically symmetric numerical simulations of the neutrino-radiation hydrodynamics of central core in massive stars. By adopting a set of progenitor models, we explore the post-bounce behavior of supernova core as well as the formation of a black hole.  
We also perform the thermal evolution of proto-neutron stars just born at the center 
through emission of supernova neutrinos.  

We demonstrate that the EOS difference is moderate in the evolution of the shock wave after the bounce and becomes notable in the late phase of proto-neutron star cooling.  
This is because the properties of hot and dense matter in the two EOS sets are similar for relatively large proton fraction at moderate density right after the core bounce and become different for neutron-rich conditions toward the formation of neutron stars.  
The timing of the formation of black holes from massive stars is found to be different due to the different maximum mass for the massive proto-neutron stars.   


This article is arranged as follows.  
We describe the EOS based on the RMF theory for numerical simulations in \S \ref{sec:eos} with 
a brief summary on modification from the Shen EOS and its consequence to the properties of neutron stars.  
After describing the setup of numerical simulations in \S \ref{sec:simulations} for core collapse of massive stars and thermal evolution of proto-neutron stars, we report the comparison of numerical results with the two sets of EOS table in post-bounce evolution, black hole formation and proto-neutron star cooling in \S \ref{sec:results}.  
We discuss the influence of the density dependence of symmetry energy in \S \ref{sec:discussion} and summarize the article in \S \ref{sec:summary}.  

\section{Equation of state for supernovae} \label{sec:eos}

\subsection{Relativistic mean field theory}\label{sec:rmf}

The Shen EOS table is constructed based on the relativistic mean field (RMF) theory and a local density approximation to describe uniform matter and non-uniform matter, respectively, in supernovae.  
The RMF theory is an effective theory of a relativistic nuclear many-body system starting from a Lagrangian, which describes nucleons interacting via the exchange of various mesons, and has been popularly used for description of nuclei and nuclear matter \citep{ser86}.  
The form of Lagrangian of RMF theory for the Shen EOS is motivated by the Dirac Br\"uckner Hatree-Fock (DBHF) theory.  
The DBHF theory is a microscopic many-body framework, which describes the nuclear saturation properties starting from the realistic nucleon-nucleon interaction, by utilizing the Dirac equation together with self-energies and the G-matrix \citep{bro90}.  
The Lagrangian includes non-linear terms of scalar and vector mesons to reproduce the value of incompressiblity at the saturation density and to describe the behavior of scalar and vector potentials in the DBHF theory.  
The parameters of the Lagrangian, TM1, are determined by the global fitting of nuclear masses and radii for stable nuclei \citep{sug94} and have been examined by the neutron skin of neutron-rich nuclei \citep{tsuz95}.  
Resulting properties of matter in neutron stars and supernovae have been reported in \citet{sum95a,sum04}.  
It is to be noted that the relativistic many-body frameworks such as the RMF theory ensure the causality of sound speed, which is often problematic in non-relativistic counterparts.  

We adopt 
the extended Lagrangian of the Shen EOS to describe the density dependence of symmetry energy by adding a coupling term of vector meson and isovector-vector meson \citep{fat10,bao14a}.  
The nucleonic Lagrangian density is composed of 
the isoscalar scalar and vector mesons ($\sigma$ and
$\omega$) and the isovector vector meson ($\rho$)
and reads
\begin{eqnarray}
\label{eq:LRMF}
\mathcal{L}_{\rm{RMF}} & = & \sum_{i=p,n}\bar{\psi}_i
\left[ i\gamma_{\mu}\partial^{\mu}-\left(M+g_{\sigma}\sigma\right)
-\gamma_{\mu} \left(g_{\omega}\omega^{\mu} +\frac{g_{\rho}}{2}
\tau_a\rho^{a\mu}\right)\right]\psi_i   \nonumber \\
&& +\frac{1}{2}\partial_{\mu}\sigma\partial^{\mu}\sigma -\frac{1}{2}%
m^2_{\sigma}\sigma^2-\frac{1}{3}g_{2}\sigma^{3} -\frac{1}{4}g_{3}\sigma^{4}
\nonumber \\
&& -\frac{1}{4}W_{\mu\nu}W^{\mu\nu} +\frac{1}{2}m^2_{\omega}\omega_{\mu}%
\omega^{\mu} +\frac{1}{4}c_{3}\left(\omega_{\mu}\omega^{\mu}\right)^2  
\nonumber \\
&& -\frac{1}{4}R^a_{\mu\nu}R^{a\mu\nu} +\frac{1}{2}m^2_{\rho}\rho^a_{\mu}%
\rho^{a\mu} +\Lambda_{\rm{v}} \left(g_{\omega}^2
\omega_{\mu}\omega^{\mu}\right)
\left(g_{\rho}^2\rho^a_{\mu}\rho^{a\mu}\right),
\end{eqnarray}
where $W^{\mu\nu}$ and $R^{a\mu\nu}$ are the antisymmetric field
tensors for $\omega^{\mu}$ and $\rho^{a\mu}$, respectively\footnote{Note that 
the coupling constant for isovector vector meson, $g_{\rho}$ is different 
by a factor of 2 from the one in \citet{sum95a}.  We follow here the convention of \citet{bao14a}.  }.
In the mean field approximation, the meson fields are treated as classical fields 
and the field operators are replaced by their expectation values. 
We derive the equations of motion for the mean fields of mesons 
and the Dirac equation for nucleons under the influence of meson fields.  
They are coupled with each other and solved in a self-consistent manner.  
For a static and uniform system, non-vanishing components are 
$\sigma =\left\langle \sigma \right\rangle$, $\omega =\left\langle
\omega^{0}\right\rangle$, and $\rho =\left\langle \rho^{30} \right\rangle$.  

For uniform matter, the energy density is given by
\begin{eqnarray}
\label{eq:ERMF}
\varepsilon &=& \displaystyle{\sum_{i=p,n} \frac{1}{\pi^2}
  \int_0^{\infty} dk\,k^2\,
  \sqrt{k^2+{M^{\ast}}^2}
  \left( f_{i+}^{k}+f_{i-}^{k}\right)
   } \nonumber\\
 & &
  +\frac{1}{2}m_{\sigma}^2\sigma^2+\frac{1}{3}g_{2}\sigma^{3}
  +\frac{1}{4}g_{3}\sigma^{4} \nonumber\\
 & &
  +\frac{1}{2}m_{\omega}^2\omega^2+\frac{3}{4}c_{3}\omega^{4}
  +\frac{1}{2}m_{\rho}^2\rho^2
  +3 \Lambda_{\rm{v}}\left(g^2_{\omega}\omega^2\right)
     \left(g^2_{\rho}\rho^2\right),
\end{eqnarray}
and the pressure is given by
\begin{eqnarray}
\label{eq:PRMF}
 P &=& \displaystyle{\sum_{i=p,n} \frac{1}{3\pi^2}
   \int_0^{\infty} dk\,k^2\,
   \frac{k^2}{\sqrt{k^2+{M^{\ast}}^2}}
   \left( f_{i+}^{k}+f_{i-}^{k}\right)
    } \nonumber\\
 & &
  -\frac{1}{2}m_{\sigma}^2\sigma^2-\frac{1}{3}g_{2}\sigma^{3}
  -\frac{1}{4}g_{3}\sigma^{4} \nonumber\\
 & &
  +\frac{1}{2}m_{\omega}^2\omega^2+\frac{1}{4}c_{3}\omega^{4}
  +\frac{1}{2}m_{\rho}^2\rho^2
  +\Lambda_{\rm{v}}\left(g^2_{\omega}\omega^2\right)
   \left(g^2_{\rho}\rho^2\right).
\end{eqnarray}
Here $M^{\ast}=M+g_{\sigma}\sigma$ is the effective nucleon mass.
$f_{i+}^{k}$ and $f_{i-}^{k}$ ($i=p,n$) are the occupation
probabilities of nucleon and antinucleon at momentum $k$.  
At zero temperature, we set $f_{i+}^{k}=1$ up to the Fermi momentum, 
which corresponds to the chemical potential 
(See also Eqs. (\ref{eq:firmf}) and (\ref{eq:nirmf}) below), 
otherwise zero and $f_{i-}^{k}=0$.  

The additional $\omega$-$\rho$ coupling term of Lagrangian effectively provides a density-dependent coupling constant between isovector meson and nucleon leading to flexibility to change the density dependence from the linear growth of symmetry energy.  
By choosing the strength of coupling constants
$g_{\rho}$ and $\Lambda_{\rm{v}}$ for isovector parts, it is possible to adjust the slope parameter of symmetry energy at the saturation density and its behavior at low and high densities.  
This modification can be also used to improve the properties of neutron matter at low density, where there are constraints from the microscopic many-body theories \citep{gan15,tew17}.  
The energy of neutron matter in TM1 tends to be small at low density and large at high density due to the linear increase of symmetry energy.  

It is advantageous that the extended RMF theory reproduces the properties of stable nuclei even after the modification of the density dependence of symmetry energy.  
By determining the parameters for the isovector interactions by fixing the symmetry energy at a density of 0.11 fm$^{-3}$, which is slightly lower than the saturation density, the binding energy of $^{208}$Pb is reproduced equally well under a wide variation of the slope parameter from $L=40$ MeV to 110.8 MeV \citep{bao14b}.  
The symmetry energy, $E_{sym}$, at the saturation density is accordingly determined by the choice of $L$.  
The thickness of neutron skin for $^{208}$Pb is shown to be dependent on the value of $L$ so that it will be examined by future experiments in a more precise manner.  

We adopt a choice of $L=40$ MeV ($E_{sym}=$31.38 MeV), which is well within the experimental constraints, for a new construction of the revised Shen EOS and explore differences from the one with $L=110.8$ MeV ($E_{sym}=$36.89 MeV).  
The EOS with $L=40$ MeV is denoted by TM1e, which is an extended version of TM1.  
These two values are the two extremes in the range studied by \citet{bao14a,bao14b,bao15} and cover a wide enough range to survey the effect of symmetry energy.  
The coupling constants for TM1 and TM1e are listed in \citet{bao14a}.  
Note that the values of $L$ and $E_{sym}$ are correlated with each other.  The choice of small (large) $L$ generally leads to a small (large) $E_{sym}$ due to the constraints on a good description of finite nuclei as seen in various sets of RMF parameters (See Table 1 of \citet{bao14b}, for example).  In the current study, the symmetry energy fixed at the density of 0.11 fm$^{-3}$ determines this correlation as described above.  Therefore, our study inevitably explores the effects of both $L$ and $E_{sym}$.  

\begin{figure}[ht!]
\plotone{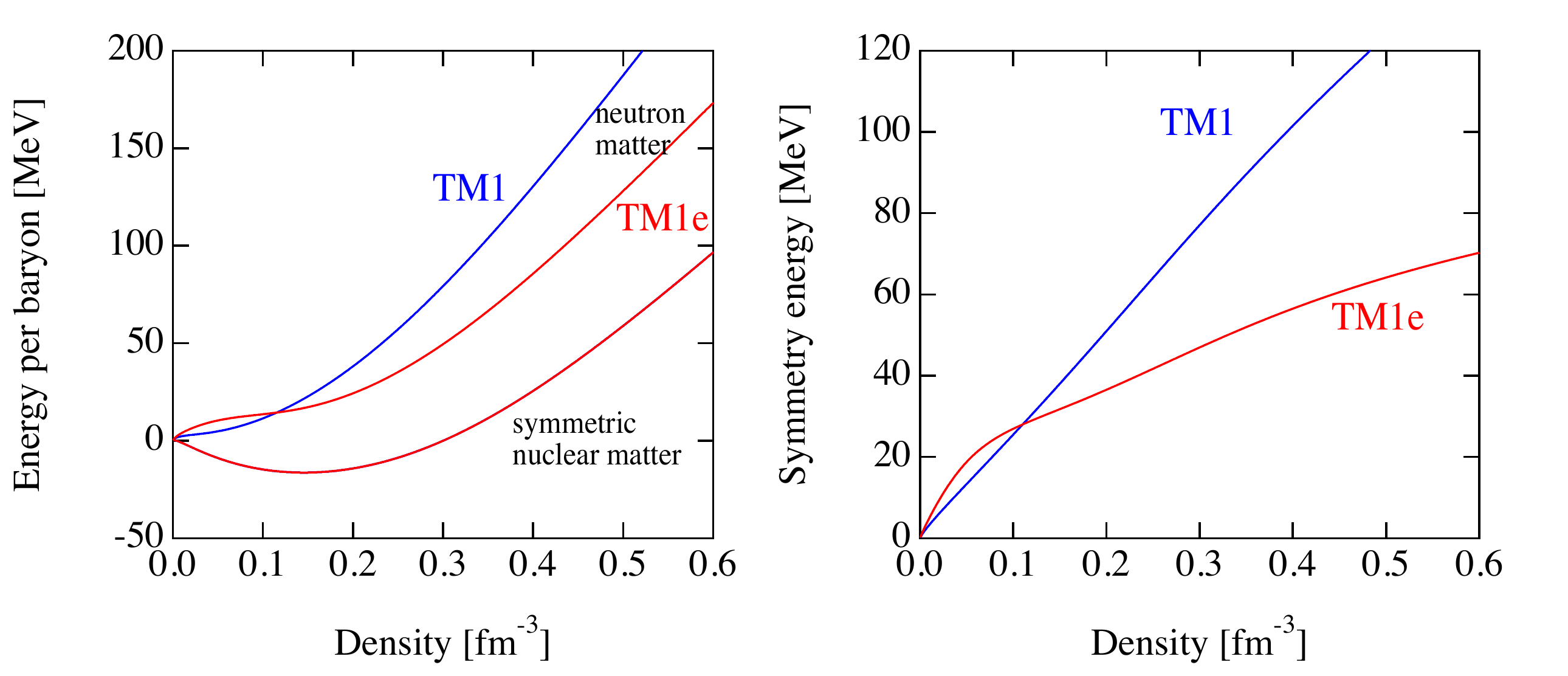}
\caption{Energy per nucleon of symmetric nuclear matter and neutron matter (left) and symmetry energy (right) as a function of nucleon density.  The results for TM1 and TM1e are shown by blue and red lines, respectively.\label{fig:EOS}}
\end{figure}

In Figure \ref{fig:EOS}, the behavior of energies of nuclear matter for the two cases of $L$ is shown as a function of nucleon density.  
At high densities, the energy per nucleon of neutron matter for $L=40$ MeV is smaller than that for $L=110.8$ MeV.  
The energy per nucleon of neutron matter for $L=40$ MeV at low densities  is slightly increased, being closer to the microscopic evaluations.  
In contrast, those of symmetric nuclear matter are the same with the common saturation point.  
The incompressibilities of symmetric matter for the two cases are both 281 MeV.  
It is convenient that one can explore modifications of neutron-rich matter under the same properties of symmetry nuclear matter so that one can take out influence of the isovector part of nuclear interaction.  
The symmetry energy for TM1e is smaller at high densities and larger at low densities with respect to those for TM1 with the linear growth.  
The resulting EOS with TM1e is softer than that with TM1 and provides different properties of neutron stars as discussed in \S \ref{sec:nstar}.  

The extension of RMF to finite temperature is done in the same way as in the Shen EOS \citep{ser86,sum94,she98a,she98b}.  
The occupation probability in the summation of states is replaced by the Fermi-Dirac distribution at the temperature and the chemical potentials for nucleons and anti-nucleons.  
The Fermi-Dirac distribution is given by
\begin{eqnarray}
\label{eq:firmf}
f_{i\pm}^{k}=\left\{1+\exp \left[ \left( \sqrt{k^{2}+{M^{\ast}}^2}
  \mp \nu_{i}\right)/T\right]
 \right\}^{-1}, 
\end{eqnarray}
where 
\begin{eqnarray}
\nu_{i} = \mu_{i} - g_{\omega}\omega - \frac{g_{\rho}}{2}\tau_i^{3}\rho 
\end{eqnarray}
and $+$ and $-$ signs denote nucleon and anti-nucleon, repectively.  
The number density of protons ($i=p$) or neutrons ($i=n$) is calculated by
\begin{equation}
\label{eq:nirmf}
 n_{i}=\frac{1}{\pi^2}
       \int_0^{\infty} dk\,k^2\,\left(f_{i+}^{k}-f_{i-}^{k}\right).
\end{equation}
The entropy density is obtained by
\begin{eqnarray}
\label{eq:SRMF}
s &=& -\displaystyle{\sum_{i=p,n}\frac{1}{\pi^{2}}
  \int_{0}^{\infty}dk\,k^{2}}
  \left[ f_{i+}^{k}\ln f_{i+}^{k}+\left( 1-f_{i+}^{k}\right)
  \ln \left(1-f_{i+}^{k}\right) \right.  \nonumber \\
& & \left. +f_{i-}^{k}\ln f_{i-}^{k}
  +\left( 1-f_{i-}^{k}\right) \ln \left( 1-f_{i-}^{k}\right) \right],
\end{eqnarray}
using the occupation probabilities.  
Note that contributions from the scalar and vector mesons are involved in the evaluation of quantities at finite temperature through the Fermi-Dirac distributions treating the effective mass and the chemical potentials (also \citet{sum95c}).  
We constructed the data table of thermodynamical quantities of uniform matter according to the format of the Shen EOS for high density part.  
The format of grid points of density, electron fraction and temperature follows the one in \citet{she11} except for the limited density range above 10$^{14}$ g/cm$^{3}$.  
The data table of Shen EOS is utilized for the low density part with the description of non-uniform matter to smoothly connect around 10$^{14.1}$ g/cm$^{3}$ to the high density part of the EOS table with TM1e.  
In this way, we can explore solely the effect of symmetry energy of uniform matter at high densities.  

\subsection{Influence of EOS on neutron stars}\label{sec:nstar}

Influence of the density-dependence of symmetry energy has been studied under the variations of $L$ from 40 MeV to 110.8 MeV in  \citet{bao14a,bao14b,bao15}.  
Regarding the properties of neutron star crusts, 
it has been shown that the neutron drip density and the composition of the crust are affected by the density-dependence of symmetry energy.  
The neutron drip density increases with the value of $L$ because of a large $L$ corresponds to a small neutron chemical potential $\mu_{n}$ at subnuclear densities.  
The crust-core transition and pasta phase are shown to be sensitive to the slope of symmetry energy.  
The density range of pasta phase is wider with more various shapes for a small $L$ as compared to the case with a large $L$ through the change in volume fraction of nuclei and in proton fraction.  

\begin{figure}[ht!]
\plotone{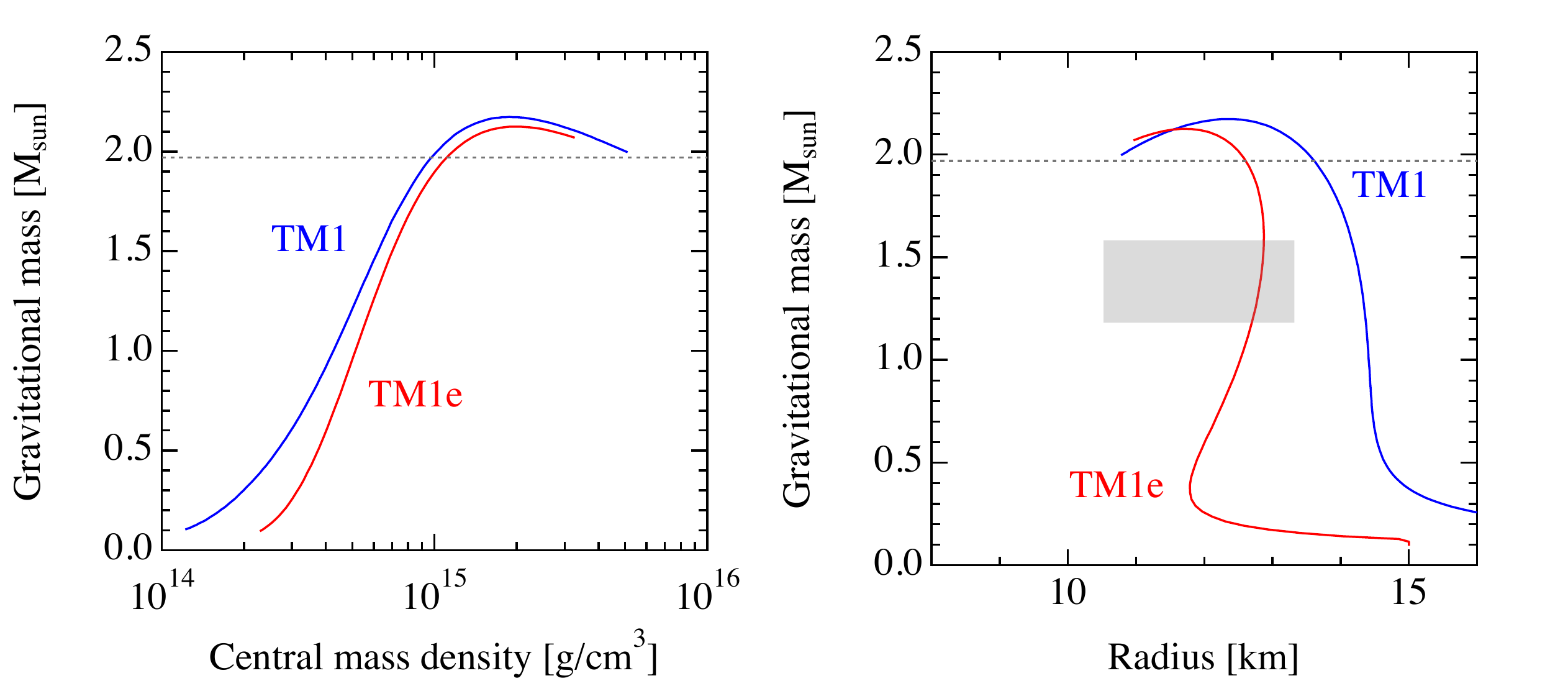}
\caption{Gravitational mass of neutron stars as a function of central mass density (left) and radius (right) for EOS with TM1 (blue) and TM1e (red).  Constraints on masses from the pulsar observation \citep{ant13} are shown by a dashed line and the range of radii derived by the gravitational wave detection is shown by the hatched area \citep{abb18}.   \label{fig:nstar}}
\end{figure}

Influence on the bulk properties of neutron star structure under the variation of $L$ is studied in the same variation of RMF  \citep{hu19}.  
Here we briefly summarize the change of neutron star properties 
for the EOSs with two $L$ values.  
Properties of neutron star matter are calculated by assuming the beta stable equilibrium and charge neutrality with the Thomas-Fermi approximation for crust region  (see \citet{hu19} for details).  
In Figure \ref{fig:nstar}, a series of neutron stars with TM1e has higher central densities than the corresponding neutron stars with TM1, reflecting a character of the softness of neutron matter seen in Fig. \ref{fig:EOS}.  
The maximum mass with TM1e is above 2M$_{\odot}$, similarly to TM1, satisfying the observational constraint on masses.  
Neutron star radii with TM1e are smaller than those with TM1.  
The newly obtained radii match well with the observational constraints in contrast to the large radii with TM1.  
The tidal deformability for TM1e is accordingly in accord with the constraint from the observation of gravitational wave  \citep{abb18,hu19}.  



\section{Setup of numerical simulations} \label{sec:simulations}

\subsection{Core collapse of massive stars} \label{sec:simccsn}

We perform spherically symmetric numerical simulations of neutrino-radiation hydrodynamics of massive stars in general relativity.  
The numerical code handles a set of equations for hydrodynamics and neutrino transfer by solving the Boltzmann equation by the S$_n$ method in a fully implicit manner.  
The data table of EOS is implemented in the code together with the neutrino reaction rates based on \citet{bru85} and its extension.  (See \citep{sum05} for details.)  
A variety of the EOS tables have been applied to explore the role of hot dense matter \citep{sum05,tog14} using this code\footnote{Application of EOS tables in the numerical code for the black hole formation can be found in \citet{sum06,sum09,nak10a,nak10b,nak12,nak13b}}.  
Some of the models studied so far with the Shen EOS are utilized as the baseline model as described below.  
In order to make simple comparisons with the previous models, we adopt the same set of neutrino reactions, which are rather simple as compared with the modern set of neutrino reactions (See \citet{pin12,len12a,fis17,kot18,nag19a}, for example).  
The four species of neutrinos are separately treated as $\nu_e$, $\bar\nu_e$, $\nu_{\mu/\tau}$ and $\bar\nu_{\mu/\tau}$.  
Here $\nu_{\mu/\tau}$ denotes the neutrinos of $\mu$- and $\tau$-types and  $\bar\nu_{\mu/\tau}$ denotes the anti-neutrinos of $\mu$- and $\tau$-types.  

Extensive studies of gravitational collapse of massive stars have been made to explore the outcome of core-collapse supernovae and the black hole formation.  
Starting with the gravitational collapse of iron core of massive stars, we follow the evolution of the shock wave after the bounce as well as the evolution of proto-neutron stars just born at the center.  
It is well known in this type of first-principle type calculations that the explosion does not occur under the spherical symmetry \citep{lie01,sum05} except for the special conditions \citep{fis11,fis18,ktak19}.  
In a certain class of massive stars, the accretion of matter intensely continues, so that a proto-neutron star becomes further massive and eventually re-collapses to the black hole \citep{lie04,sum06,ocon11}.  
The numerical code is utilized to reveal also the properties of neutrino emission during the evolution and provides the basic data for observation of supernova neutrino bursts \citep{nak13a,suw19} and diffuse supernova neutrino background \citep{nak13c,nak15,yuk15,hor18}.  

We explore the influence of the symmetry energy in the evolution from the gravitational collapse of massive stars of 15M$_{\odot}$ and 40M$_{\odot}$ by \citet{woo95}.  
We explore also the cases with massive stars of 11.2M$_{\odot}$ and 15M$_{\odot}$ \citep{woo02} as well as a massive star of 50M$_{\odot}$ by \citet{tom07} to explore whether the influence is general.  
The massive stars of 40M$_{\odot}$ and 50M$_{\odot}$ are adopted to examine the case of black hole formations.  
We newly performed numerical simulations with the EOS of TM1e for all models listed above.  

We reanalyze the results of numerical simulations with the EOS of TM1 for massive stars of 15M$_{\odot}$ and 40M$_{\odot}$ by \citet{woo95} and 50M$_{\odot}$ by \citet{tom07}, which are reported in the articles \citep{sum05,sum07,sum08}.  
We additionally performed numerical simulations with the EOS of TM1 for massive stars of 11.2M$_{\odot}$ and 15M$_{\odot}$ \citep{woo02}.  

For comparison, we adopt 255 nonuniform spatial zones in the mass coordinate for the radial direction with 6 angle zones and 14 energy zones as the default resolution that has been adopted for previous simulations.  
We utilize the regrid method to resolve the accretion of matter in the case of black hole formation.  
In the new models of black hole formation, we adopt 511 nonuniform spatial zones by remapping of grids to capture the slow accretion of matter in the late stage at $\sim$0.5 s after bounce.  

\subsection{Proto-neutron star cooling} \label{sec:simpnsc}

We perform spherically symmetric numerical simulations of the thermal evolution of proto-neutron stars in general relativity.  
The numerical code solves a set of equations for quasi-hydrostatic structure of proto-neutron stars together with thermal and compositional evolution through neutrino transfer by the flux-limited diffusion approximation \citep{suz94,nak13a}.  
We follow the cooling of proto-neutron stars through neutrino emission starting from the configuration soon after the core bounce.  
The data table of EOS is implemented in the code with a basic set of neutrino reactions, which corresponds to the reaction set used in the core-collapse simulations described above in \S \ref{sec:simccsn}.  
The three species of neutrinos are separately treated as $\nu_e$, $\bar\nu_e$ and $\nu_{x}$, where $\nu_{\mu}$, $\bar\nu_{\mu}$, $\nu_{\tau}$ and $\bar\nu_{\tau}$ are collectively treated as $\nu_{x}$.  
This code has been used to provide the supernova neutrino database \citep{nak13a,suw19} and to explore the influence of EOS \citep{sum95c,nak18,nak19} and neutrino reactions \citep{suz93}.  

The initial model is given by fixing the distribution of the entropy $S$ and electron fraction $Y_e$ as a function of the mass coordinate taken from a numerical simulation of core-collapse supernovae in the same way as in \citet{nak18}.  
The initial $S$ and $Y_e$ distributions are taken from the profile at 0.3 s after the core bounce of the 15M$_{\odot}$ star with the Shen EOS\footnote{These $S$ and $Y_e$ profiles are identical to those for the initial model with the Shen EOS in \citet{nak18}.  }.  
We set up initial neutrino distributions through relaxing evolution so that the steady flow of neutrinos is achieved by fixing $S$ and $Y_e$ distributions under the hydrostatic structure for each case of EOS table.  
This profile is used as a typical situation for the nascent proto-neutron star.  
In principle, the initial profiles should be taken from the core-collapse simulations covering from the collapse of massive stars to the successful explosion for the rigorous comparison of various sets of EOS.  
However, the consistently connecting simulation is not simple because of non-explosion under the spherical symmetry.  
Hence, we fix the common profile for the two EOS cases and examine the influence after the birth.  
In fact, as we will show in \S \ref{sec:outccsn}, the proto-neutron stars born at the center are quite similar to each other for the TM1 and TM1e EOS.  
Hence, the start of the common initial profiles for proto-neutron star evolution is permitted as long as we explore the long-term effect of evolution over 10 s.  

\section{Numerical results} \label{sec:results}

\subsection{Post bounce evolution} \label{sec:outccsn}

\begin{figure}[ht!]
\epsscale{0.5}
\plotone{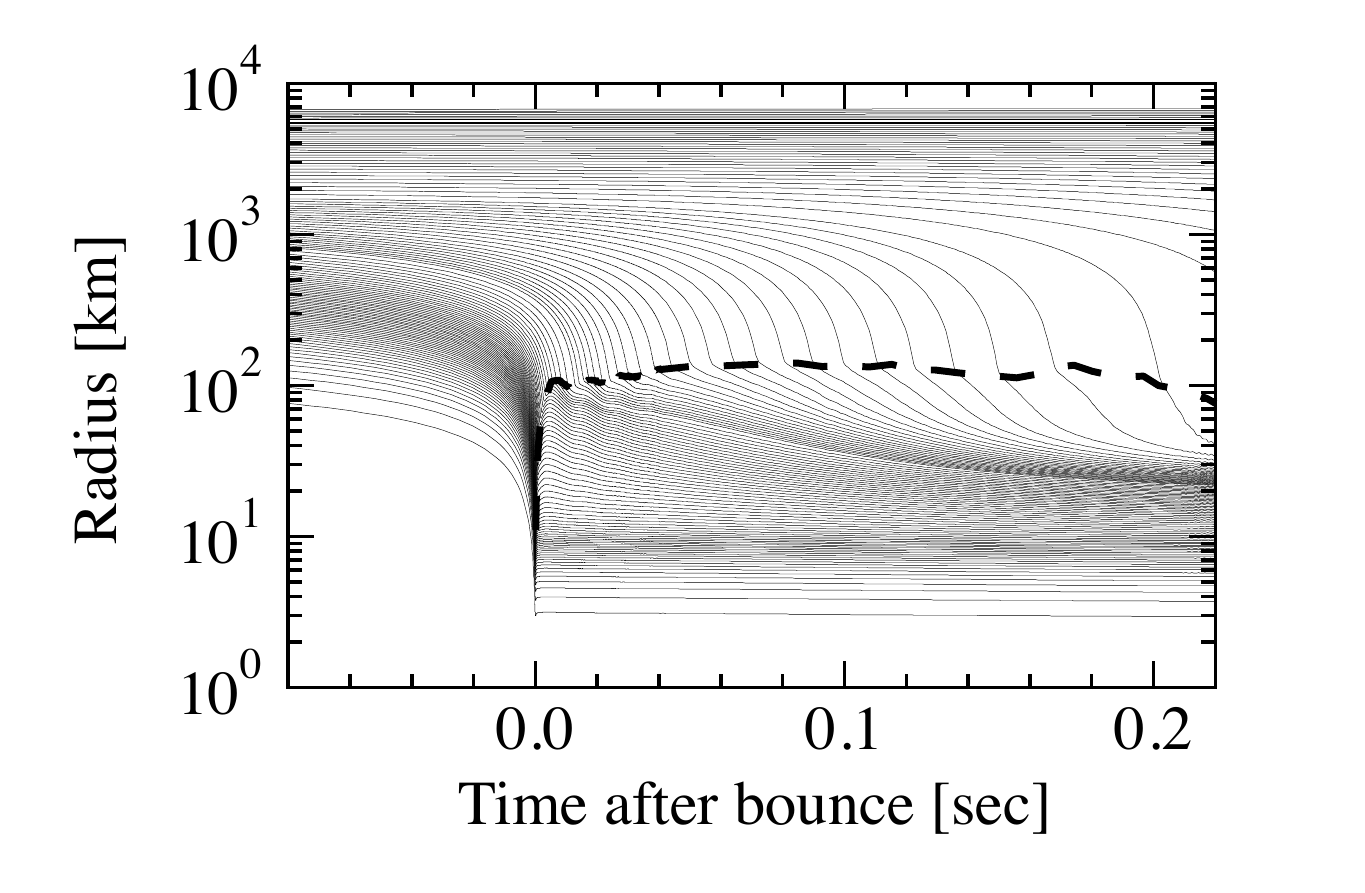}
\caption{Trajectories of gravitational collapse and core bounce of central core of the 15M$_{\odot}$ star with the TM1e EOS.  
Radial positions of fluid elements at baryon mass coordinate are shown as a function of the time after bounce.  
The radial positions are plotted for 0.02M$_{\odot}$ and  0.01M$_{\odot}$ spacing inside and outside 1.2M$_{\odot}$, respectively, in the mass coordinate.  
Position of shock wave is also plotted by dashed line.  \label{fig:15Mtraj}}
\end{figure}

We start with the numerical results for the gravitational collapse and core bounce of the 15M$_{\odot}$ star by \citet{woo95} using the TM1e EOS in conjunction with the TM1 EOS.  
In Figure \ref{fig:15Mtraj}, radial trajectories of fluid elements are plotted with the position of shock wave as a function of time after the core bounce.  
The shock wave is launched beyond 100 km and stalls around there without any sign of the revival for explosion.  
The evolution of shock wave is nearly the same as in the case with the TM1 EOS \citep{sum05}.  
Note that the evolution of the gravitational collapse up to the point when the central density exceeds over 10$^{14}$ g/cm$^{3}$ is common and the EOS is switched to the TM1e.  
Therefore, we solely examine the influence of different behavior of the symmetry energy in uniform matter.  

\begin{figure}[ht!]
\plotone{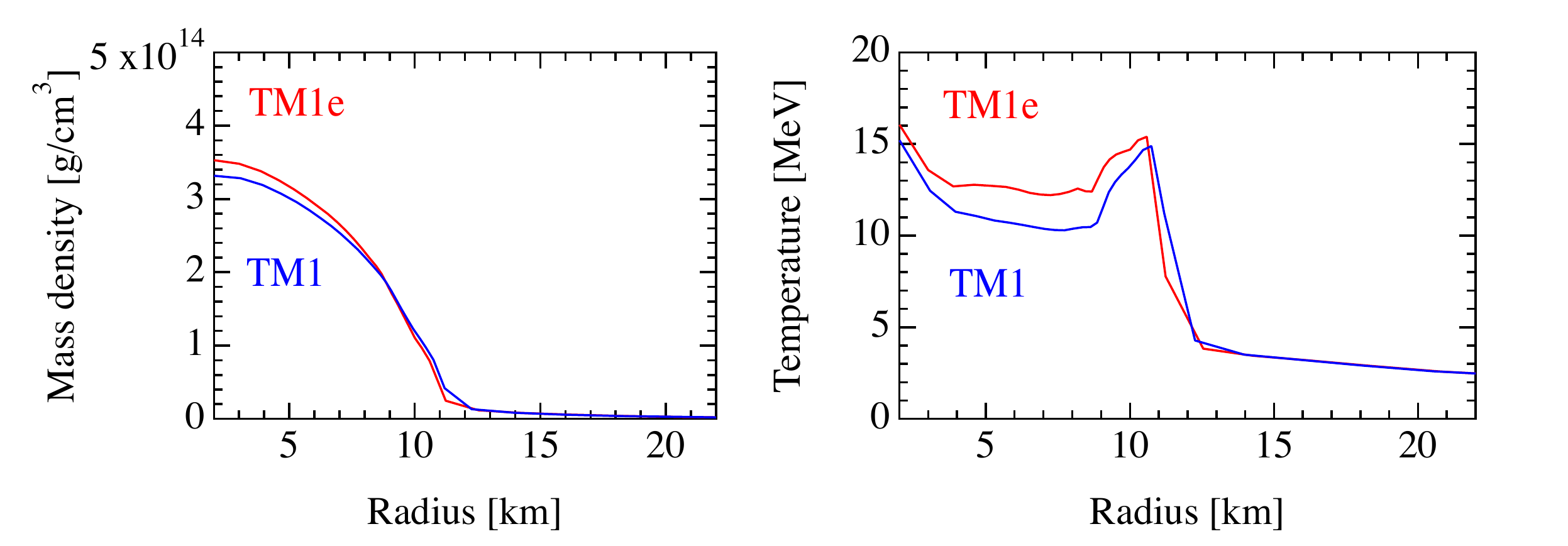}
\caption{Profiles of mass density (left) and temperature (right) at the core bounce are shown as a function of radius for the models of the 15M$_{\odot}$ star with the TM1 (blue) and TM1e (red) EOSs.   \label{fig:15M0msRhoT}} 
\end{figure}

\begin{figure}[ht!]
\plotone{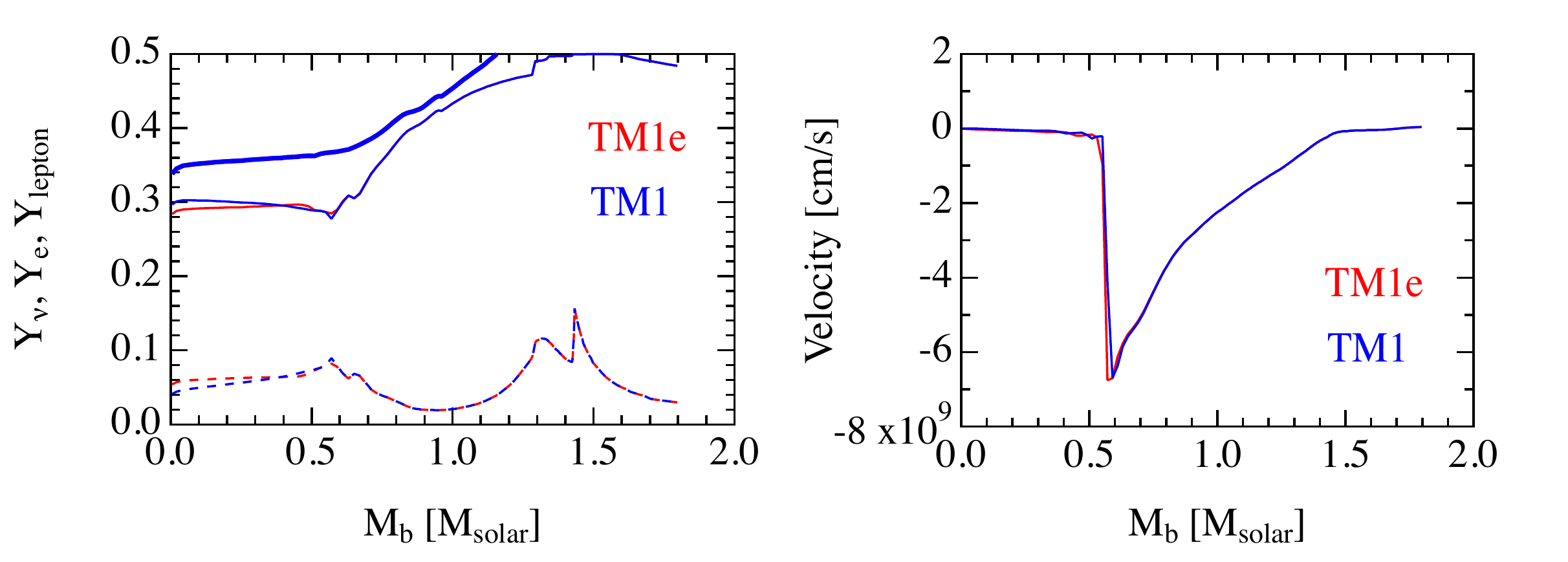}
\caption{Profiles of lepton fractions (left) and velocity (right) at the core bounce are shown as a function of baryon mass coordinate for the models of the 15M$_{\odot}$ star with the TM1 (blue) and TM1e (red) EOSs.  
In the left panel, the thick solid, solid and dashed lines denote the lepton, electron and electron-type neutrino fractions, respectively.  \label{fig:15M0msYeV}}
\end{figure}

The difference between the TM1 and TM1e EOSs is small at the core bounce and appears in the limited central region.  
The central density for the TM1e is slightly higher than that for the TM1 due to the softness.  
Accordingly the temperature for the TM1e is higher as seen in Fig.  \ref{fig:15M0msRhoT}.  
The lepton fraction is the same for the two cases as in Fig. \ref{fig:15M0msYeV} because of the common evolution of gravitational collapse.  
The electron fraction in the central region with the TM1e is slightly smaller because the smaller symmetry energy at high densities allows a smaller proton fraction.  
The size of the bounce core remains the same having the same fraction of trapped leptons.  
The positions of the shock wave at the core bounce are almost the same as can be seen in Fig. \ref{fig:15M0msYeV} and in the position of temperature peak in Fig. \ref{fig:15M0msRhoT}.  
Differences of the core bounce turn out to be small because the density is not so high and the proton fraction is still around 0.3.  
The nuclear matter just above the saturation density is not so neutron-rich as to have large differences.  
Hence, the initial situations of shock wave for different symmetry energies are similar as long as the uniform matter at high densities is concerned.  
It should be noted that effects of symmetry energy can appear during the gravitational collapse through electron captures on protons and nuclei.  

\begin{figure}[ht!]
\plotone{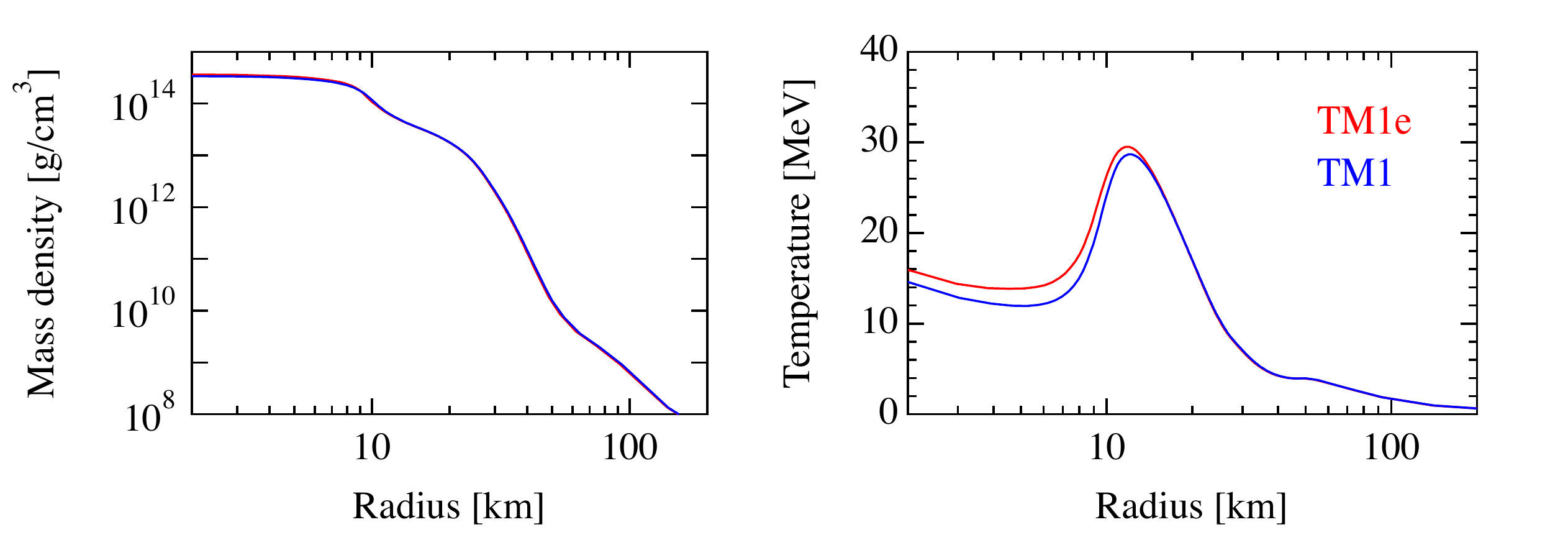}
\plotone{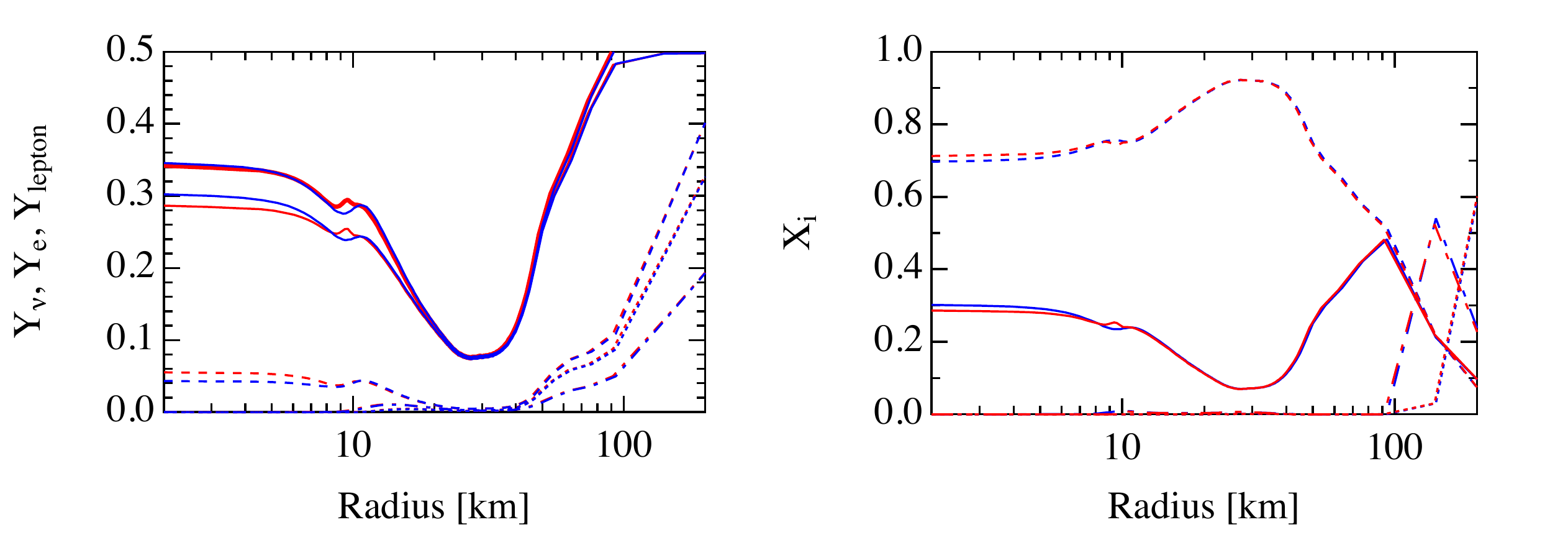}
\caption{Profiles of mass density (top left), temperature (top right), lepton fractions (bottom left) and mass fractions (bottom right) at 200 ms after the core bounce are shown as a function of radius for the models of the 15M$_{\odot}$ star with the TM1 (blue) and TM1e (red) EOSs. In the bottom left panel, thick solid, solid, dashed, dotted and dash-dotted lines display the lepton, electron, electron-type neutrino, electron-type anti-neutrino, $\mu$-type neutrino fractions, respectively.  In the bottom right panel, solid, dashed, dotted and dash-dotted lines display the mass fraction of proton, neutron, nuclei and alpha-particles, respectively.   \label{fig:15M200ms}}
\end{figure}

The tendency of small differences continues for several hundred milliseconds after the core bounce.  
In Figure \ref{fig:15M200ms}, we show the profiles of the central core of the 15M$_{\odot}$ star at 200 ms after core bounce.  
Density profiles are similar with a slight difference of central densities, 3.4$\times10^{14}$ g/cm$^{3}$ and 3.6$\times10^{14}$ g/cm$^{3}$ for the TM1 and TM1e cases, respectively.  
The temperature for TM1e is slightly higher than that for TM1 in the central part of a nascent proto-neutron star just born.  
Profiles of lepton fraction are similar for two cases with a slight difference for proton fractions due to the symmetry energy effect mentioned above.  
The compositions inside the proto-neutron star are similar with a slight difference in proton and neutron fractions accordingly.  

\begin{figure}[ht!]
\plotone{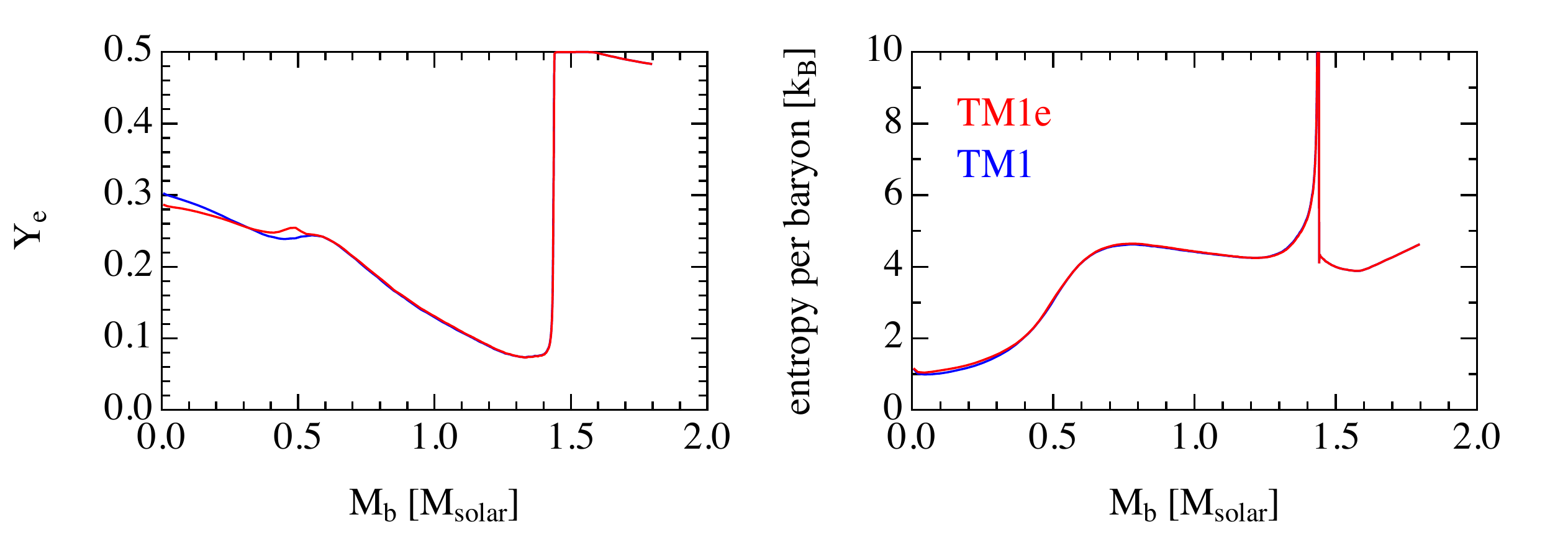}
\caption{Profiles of electron fraction (left) and entropy per baryon (right) at 200 ms after the core bounce are shown as a function of baryon mass coordinate for the models of the 15M$_{\odot}$ star with the TM1 (blue) and TM1e (red) EOSs.  \label{fig:15M200msSYe}}
\end{figure}

In the current study for the influence of uniform matter, the profiles of proto-neutron stars are similar at the initial stage.  
In Fig. \ref{fig:15M200msSYe}, we show the profiles of the electron fraction and entropy per baryon at 200 ms after the bounce as a function of baryon mass coordinate.  
The two profiles are similar except for slight differences inside $\sim$0.5M$_{\odot}$.  
This fact supports our usage of the common initial condition for proto-neutron star cooling in \S \ref{sec:outpnsc}.  

\begin{figure}[ht!]
\plotone{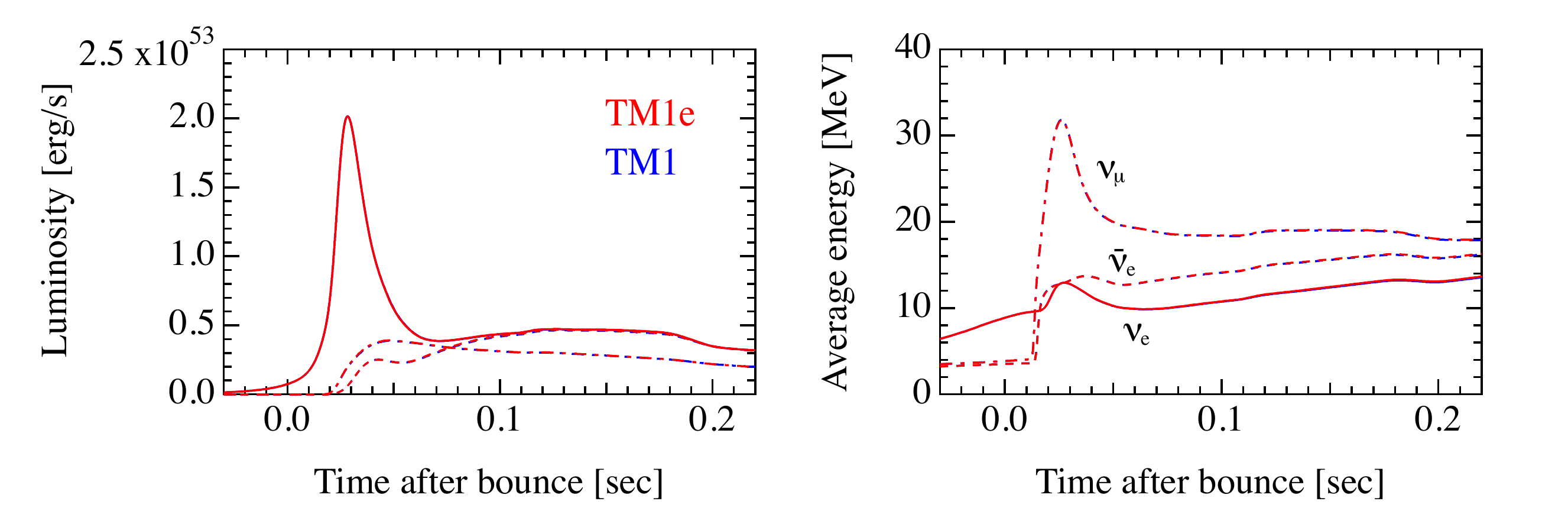}
\caption{Evolution of luminosities and average energies of neutrinos emitted during the collapse and bounce of 15M$_{\odot}$ star with TM1 (blue) and TM1e (red) as a function of time after bounce.  Quantities for the three neutrino species, $\nu_e$, $\bar\nu_e$, $\nu_{\mu}$ are shown by solid, dashed and dash-dotted lines, respectively.  \label{fig:15Mlnuenu}}
\end{figure}

Because of the similar evolution of central cores, emission of neutrinos during the collapse and post-bounce proceeds in nearly the same way for both cases.  
Figure \ref{fig:15Mlnuenu} displays the luminosities and average energies of neutrinos emitted from the central core of the 15M$_{\odot}$ star.  
The neutrino signals for the two cases are hardly discernible in the early phase for $\sim$200 ms after the core bounce.  
This is because we follow the same evolution during the collapse and the outer profiles for neutrino emissions at $\sim10^{11}$ g/cm$^{3}$ are the same even for a while after bounce.  
It would take a long time over 1 s to see the difference through the diffusion of neutrinos inside the central region.  
In the evolution of proto-neutron stars (\S \ref{sec:outpnsc}), the influence will be more enhanced when the matter becomes denser and more neutron-rich after the deleptonization.  

\begin{figure}[ht!]
\plotone{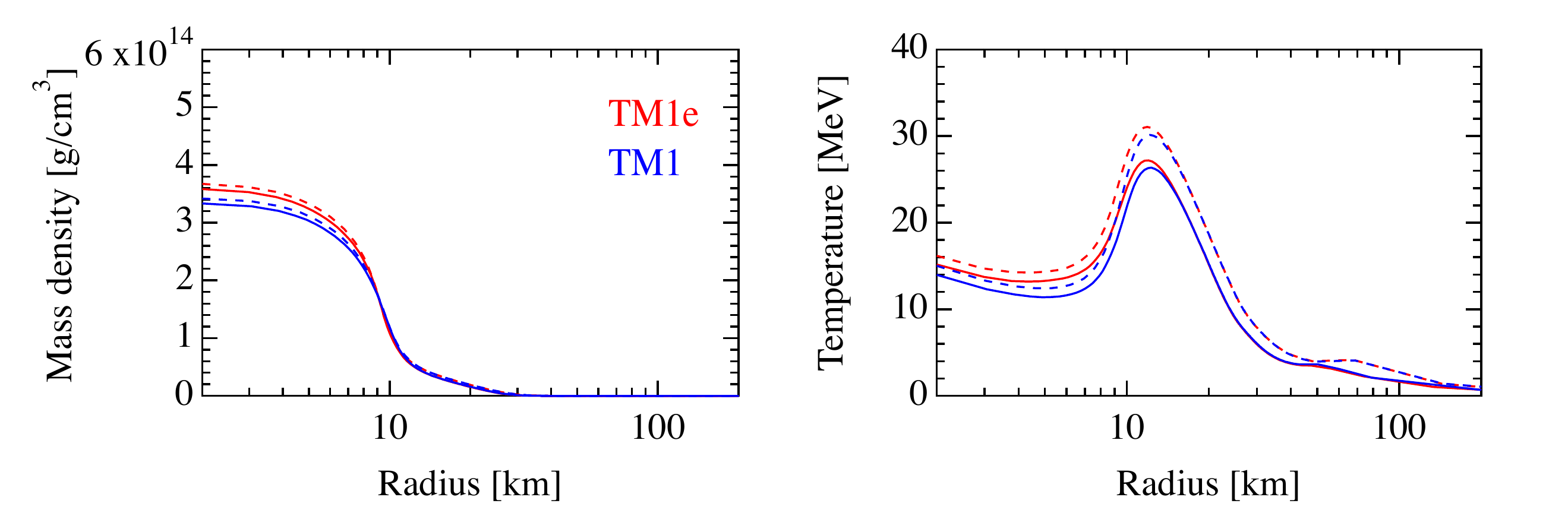}
\caption{Profiles of density (left) and temperature (right) at 200 ms after bounce as a function of radius for the case of 11.2M$_{\odot}$ (solid line) and 15M$_{\odot}$ (dashed line) stars \citep{woo02} using the TM1 (blue) and TM1e (red) EOS.  \label{fig:11M200ms}}
\end{figure}

The effects of the symmetry energy are seen in the same way for additional models of massive stars.  
We examined the numerical results for the 11.2M$_{\odot}$ and 15M$_{\odot}$ stars by \citet{woo02} to check the similarity and difference using the TM1 and TM1e EOSs.  
We found that the the overall dynamics from the core bounce to the stall of shock propagation are similar to each other between the two EOS cases.  
The profiles at the core bounce have only small differences, as in the case of 15M$_{\odot}$ star, having larger densities and temperatures and smaller electron fractions at center in TM1e.  
The difference of temperature in the central part is seen modestly among them, but remains less than $\sim$2 MeV.  
We show in Fig. \ref{fig:11M200ms} the profiles of density and temperature at 200 ms after bounce for the 11.2M$_{\odot}$ and 15M$_{\odot}$ stars with two EOSs.  
The central density and temperature for TM1e are higher than those for TM1 for the two massive star models.  
These differences are confined inside 10 km where the density is higher than 10$^{14}$ g/cm$^{3}$.  
Note that the baryon mass of central object (i.e. newly born proto-neutron star) varies depending on the profile of massive stars through the accretion rate.  
The baryon mass inside 10$^{11}$ g/cm$^{3}$ is 1.31M$_{\odot}$ and 1.53M$_{\odot}$ in the  11.2M$_{\odot}$ and 15M$_{\odot}$ stars
\footnote{ The corresponding value for the 15M$_{\odot}$ star by \citet{woo95} is 1.43M$_{\odot}$.}, respectively, for both EOSs.  
The central density and temperature for the 15M$_{\odot}$ star is accordingly higher.  
The difference due to the symmetry energy may turn out to be comparable or even less as compared with that due to the initial model of massive stars.  
The properties of neutrino emission are almost the same between the two EOSs up to 200 ms after bounce.  
It is to be noted that the neutrino signals are different for different progenitors 
and the differences from EOS effects are much smaller.   

\subsection{Black hole formation} \label{sec:outbh}

\begin{figure}[ht!]
\epsscale{0.5}
\plotone{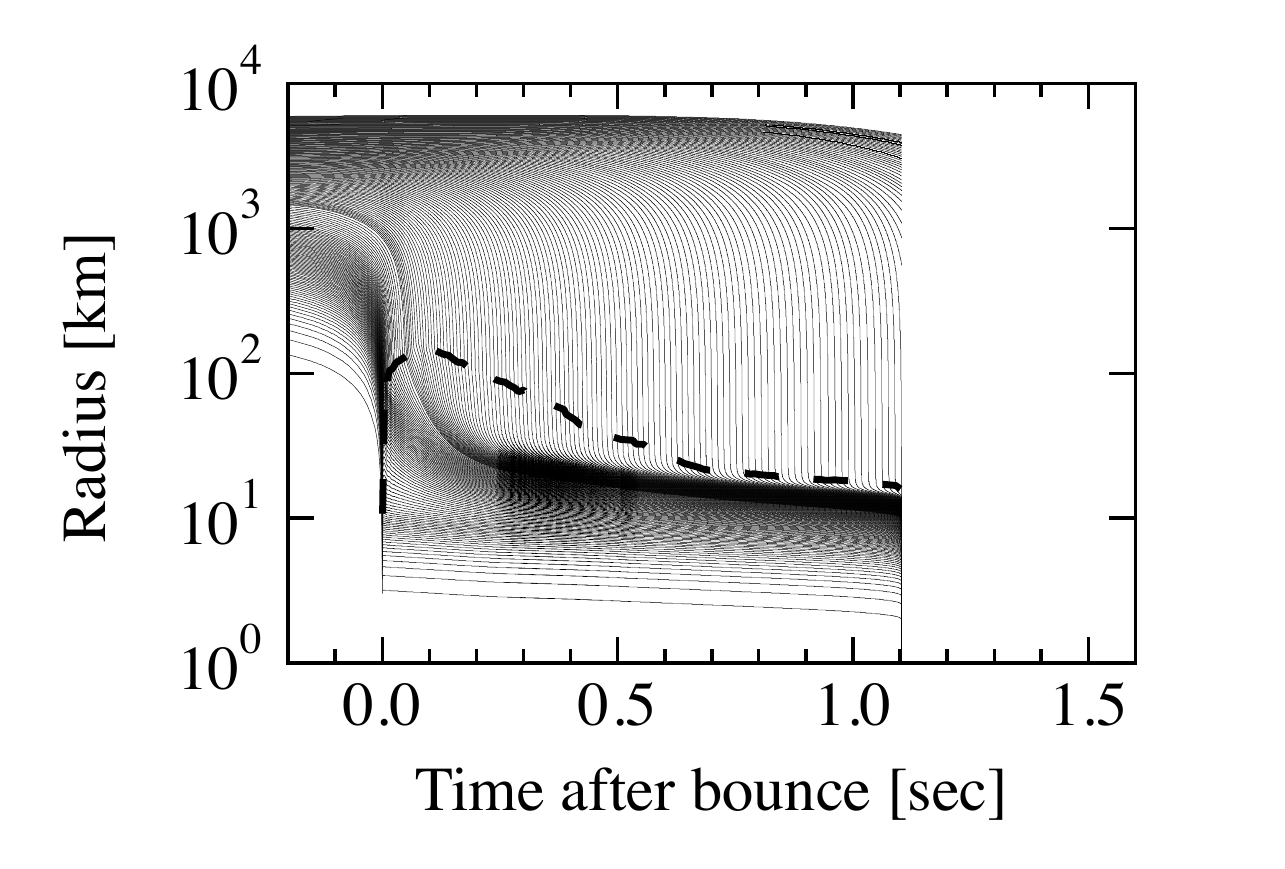}
\plotone{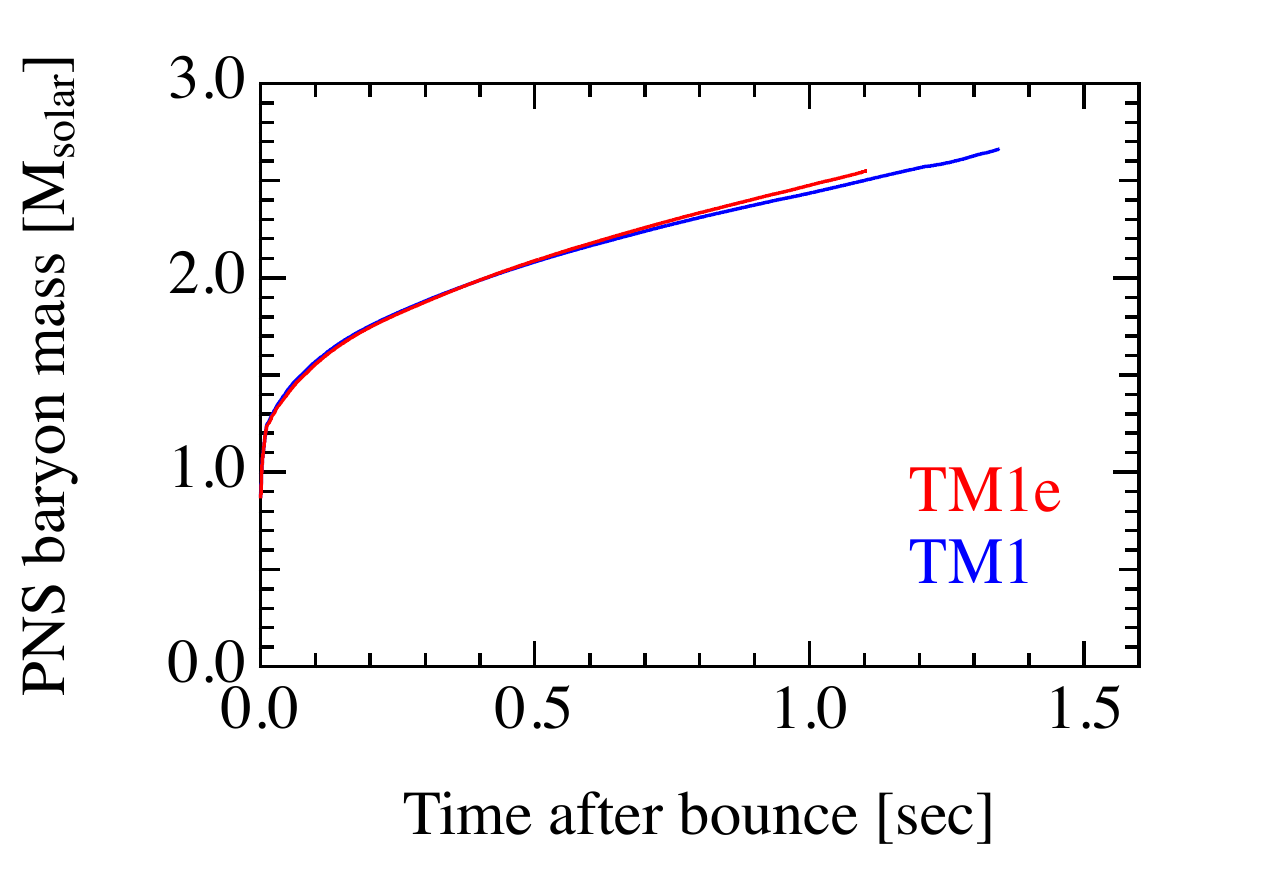}
\caption{(left) Trajectories of gravitational collapse, core bounce and re-collapse of central core of 40M$_{\odot}$ star with the TM1e EOS.  
Radial positions of fluid elements at the baryon mass coordinate are shown as a function of time after bounce.  
The radial positions are plotted for 0.02M$_{\odot}$ and  0.01M$_{\odot}$ spacing inside and outside 1.5M$_{\odot}$, respectively, in the mass coordinate.  
Position of shock wave is also plotted by a dashed line.  
(right) Baryon mass of proto-neutron stars for the TM1 (blue) and TM1e (red) cases as a function of time after bounce.  \label{fig:40Mtraj}}
\end{figure}

We first describe the numerical results for core-collapse and post-bounce evolution from the 40M$_{\odot}$ star by \citet{woo95} to explore the influence of the symmetry energy on the black hole formation.  
In Figure \ref{fig:40Mtraj}, we show the radial trajectories of fluid elements for the case with TM1e and the baryon mass of proto-neutron star for the two cases as a function of time after bounce.  
The general behavior is similar to the case with TM1 except for a shorter duration until the black hole formation.  
After the core collapse and bounce, the shock wave stalls due to the intense accretion of matter and the central object becomes increasingly massive.  
The proto-neutron star mass, which is defined here as the baryon mass inside the radius at the density of 10$^{11}$ g/cm$^{3}$, increases over 2M$_{\odot}$ at the accretion rate of $\dot{{\rm M}}\sim$0.7M$_{\odot}$/s.  
In TM1e, the central object collapses to a black hole at 1.1 s after bounce, which is $\sim$0.2 s shorter than the TM1 case.  
The increasing evolution of the proto-neutron star mass is common for the two EOSs since the accretion rate is mostly determined by the density profile of massive star.  
The shorter duration is solely determined by the critical mass of hot and lepton-rich proto-neutron stars, which is in turn determined by the EOS.  
The smaller maximum mass by the TM1e EOS leads to the shorter duration than that brought by the TM1 EOS.  
Note that the critical mass of proto-neutron stars depends on the lepton fractions and entropy and is different from that of cold neutron stars.  

\begin{figure}[ht!]
\plotone{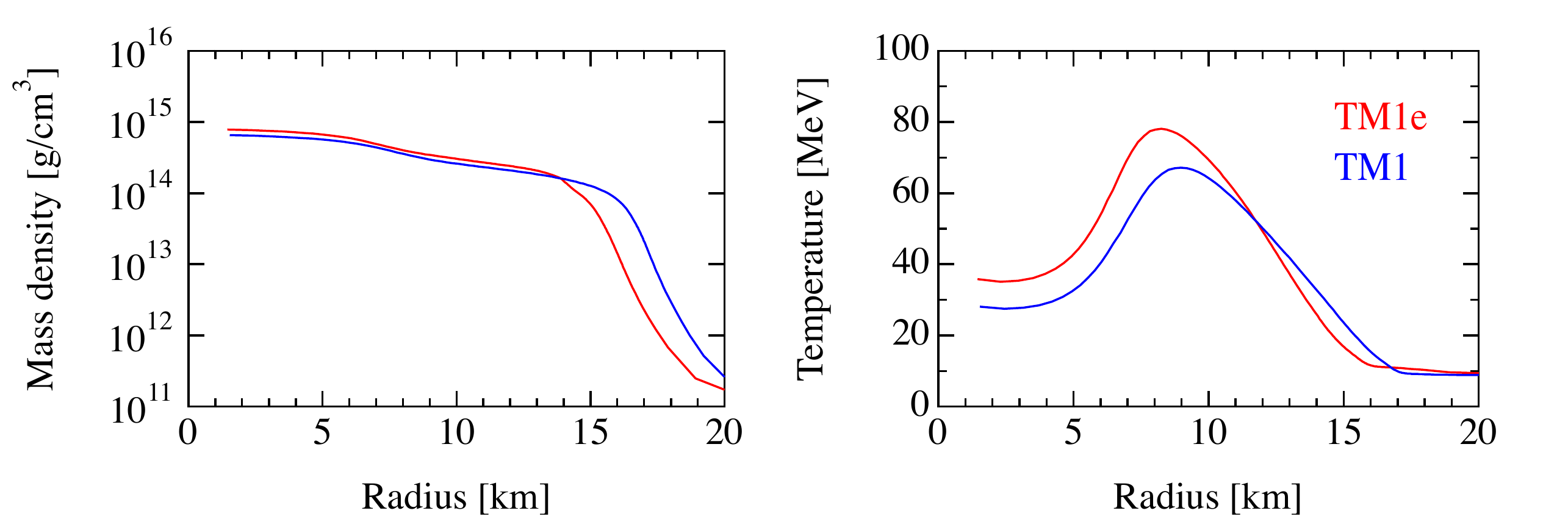}
\plotone{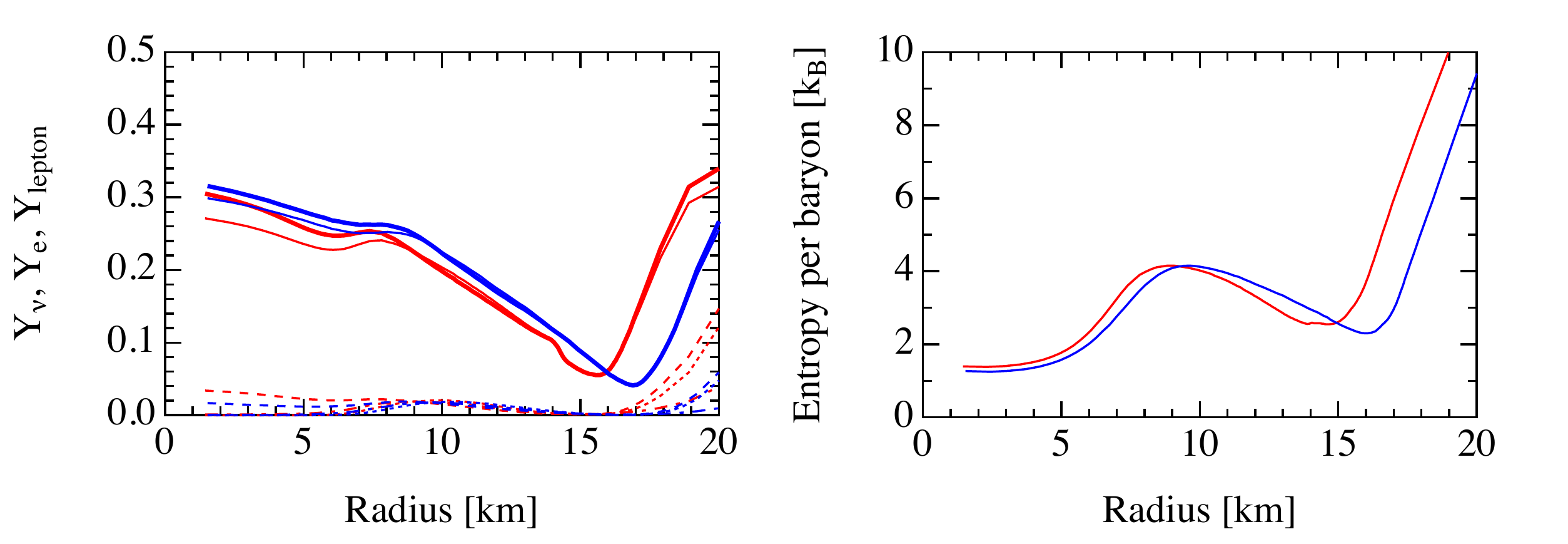}
\caption{Profiles of density (top left), temperature (top right), electron and lepton fractions (bottom left) and entropy per baryon (bottom right) of 40M$_{\odot}$ star at 1.0 s after bounce with TM1 (blue) and TM1e (red) as a function of radius. 
In the bottom left panel, thick solid, solid, dashed, dotted and dash-dotted lines display the lepton, electron, electron-type neutrino, electron-type anti-neutrino, $\mu$-type neutrino fractions, respectively.  \label{fig:40Mt1RhoT}}
\end{figure}

Figure \ref{fig:40Mt1RhoT} shows the properties of massive proto-neutron stars at 1.0 s after bounce for TM1 and TM1e models.  
The proto-neutron star of TM1e is  more compact and hotter than that of TM1.  
Distributions of lepton fraction and entropy per baryon, which characterize the properties of dense matter, are similar to each other except for radial shift.  
The electron fraction for TM1e is slightly smaller than that for TM1 because of the small symmetry energy.  
This difference of $Y_e$ appears at the core bounce and remains until the black hole formation.  
The difference of temperature also starts from the core bounce and becomes large at later stages.  
This causes the difference of neutrino average energies as will be shown below.  

\begin{figure}[ht!]
\plotone{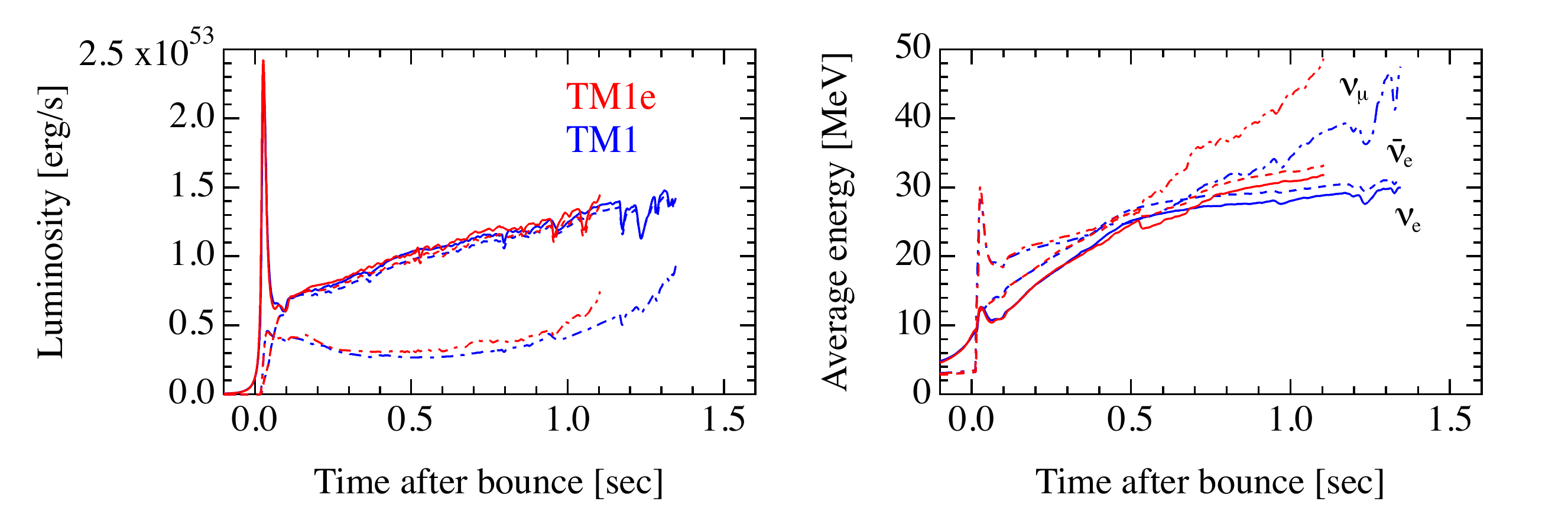}
\caption{Evolution of luminosities (left) and average energies (right) of neutrinos emitted during the core bounce toward the black hole formation of 40M$_{\odot}$ star with TM1 (blue) and TM1e (red) as a function of time after bounce.  Quantities for the three neutrino species, $\nu_e$, $\bar\nu_e$, $\nu_{\mu}$ are shown by solid, dashed and dash-dotted lines, respectively.  Note that there is a connection of lines due to the rezoning made at $\sim0.5$ s after bounce for TM1e.  \label{fig:40Mlnuenu}}
\end{figure}

We show in Fig. \ref{fig:40Mlnuenu} the time evolution of luminosities and average energies of neutrinos emitted toward the black hole formation.  
The neutrino burst lasts for $\sim$1 s and is terminated at the black hole formation.  
This character of short duration can be used to identify the event of black hole formation from massive stars in principle.  
The average energies increase due to the rising temperature inside the proto-neutron stars.  
After the neutronization burst similar to ordinary supernovae, the electron-type neutrinos and anti-neutrinos are mainly originated from the accretion of matter while the muon-type (anti-)neutrinos diffuse out from the hot central object.  
The duration of neutrino burst is shorter in TM1e than that in TM1 due to the earlier collapse to the black hole.  
The end point of burst is determined by the critical mass of proto-neutron star as discussed above and has dependence on the EOS, i.e. the symmetry energy in this case.  

\begin{figure}[ht!]
\plotone{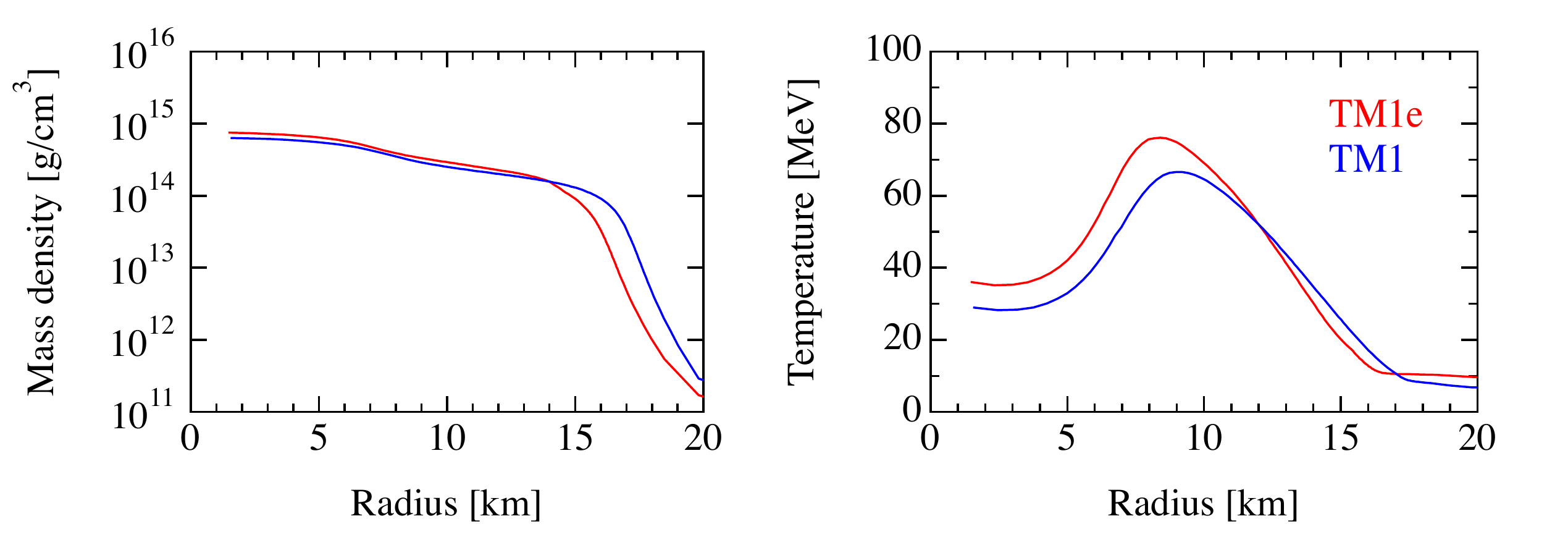}
\caption{Profiles of density (left) and temperature (right) at 1 s after bounce are shown a function of radius for the case of 50M$_{\odot}$ stars \citep{tom07} using the TM1 (blue) and TM1e (red) EOS.  
\label{fig:50Mt1RhoT}}
\end{figure}

\begin{figure}[ht!]
\epsscale{0.5}
\plotone{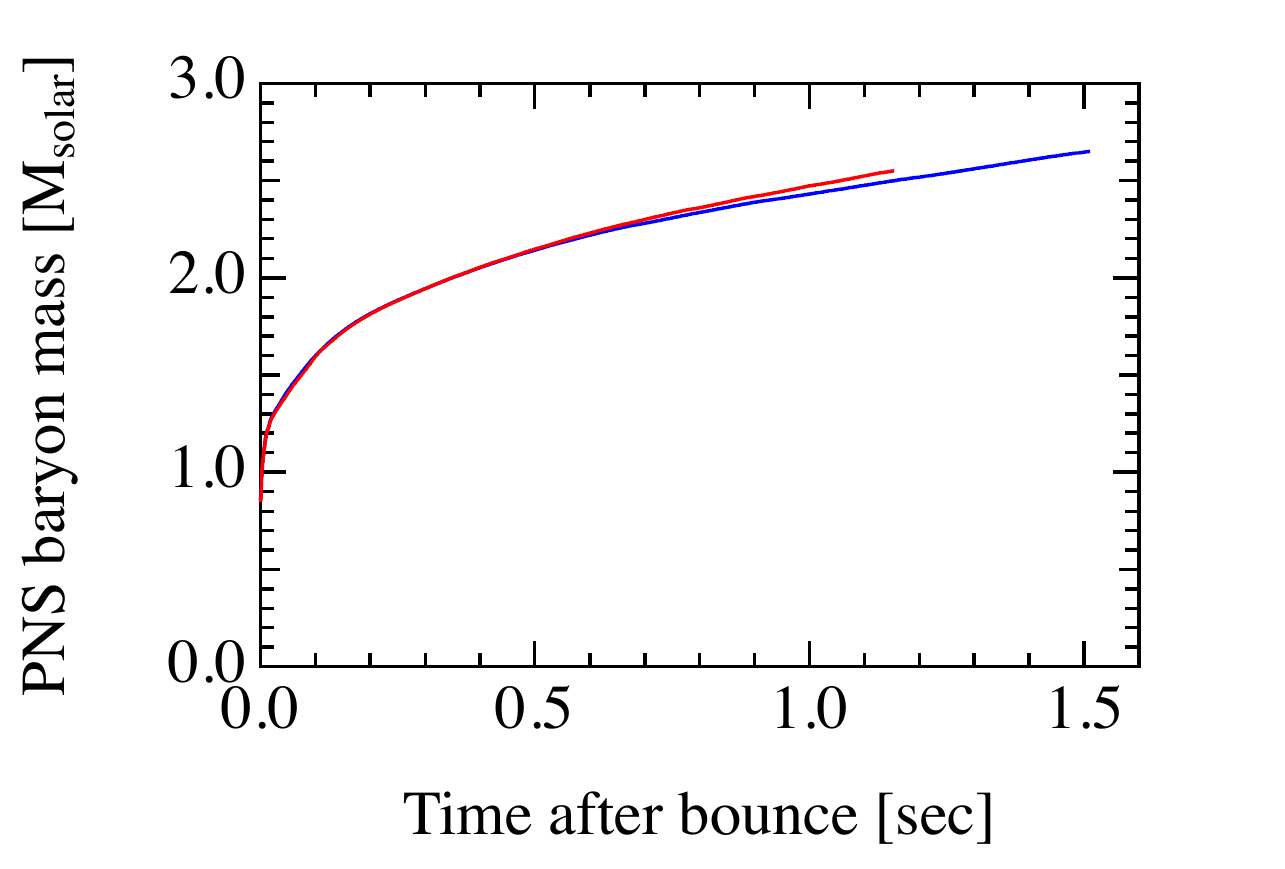}
\plotone{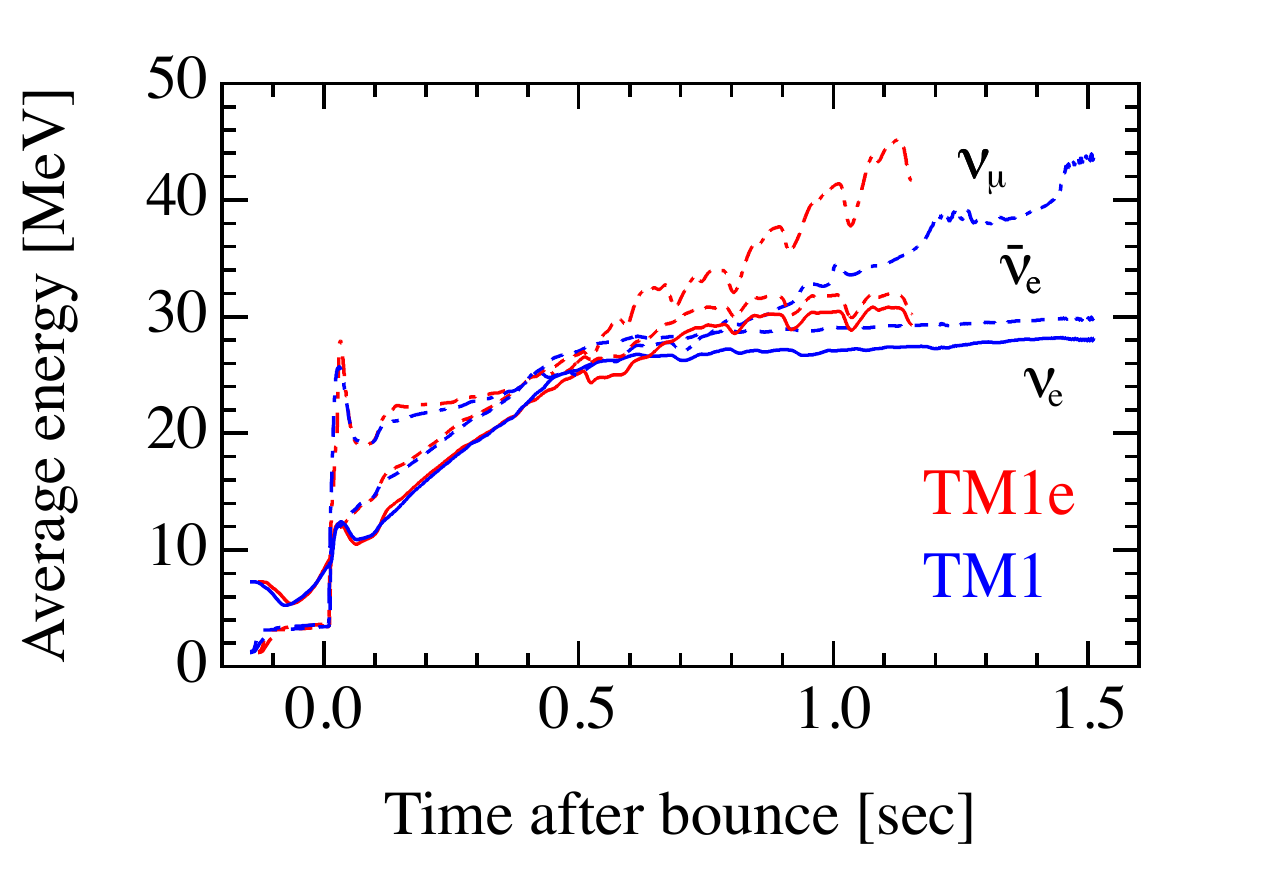}
\caption{Evolution of the baryon mass of proto-neutron stars (left) and the average energies of neutrinos (right) emitted during the core bounce toward the black hole formation of 50M$_{\odot}$ star with TM1 (blue) and TM1e (red) as a function of time after bounce.  Quantities for the three neutrino species, $\nu_e$, $\bar\nu_e$, $\nu_{\mu}$ are shown by solid, dashed and dash-dotted lines, respectively.  Note that there is a connection of lines due to the rezoning made at $\sim0.5$ s after bounce for TM1e.  
\label{fig:50MMpnsEnu}}
\end{figure}

The different duration of neutrino burst seen above can be a probe of the EOS, if the initial profile of central core in massive stars is given.  
However, the duration in reality depends on unknown profile of the progenitor \citep{sum08}.  
As an example in different progenitors, we examined the numerical results for the 50M$_{\odot}$ star by \citet{tom07} to check the general tendency of the EOS influence.  
We found that the behavior of evolution toward the black hole is common, but the duration is shorter for TM1e than TM1 also in this case.  
We see different profiles of density and temperature in the same way as the case for 40M$_{\odot}$ star.  
Figure \ref{fig:50Mt1RhoT} shows the profiles of density and temperature at 1 s after bounce for the two EOSs.  
The massive proto-neutron star for TM1e is more compact and dense having higher temperature as compared with TM1.  
Note that the total baryon mass of proto-neutron star is almost the same at this time snapshot as shown in Fig. \ref{fig:50MMpnsEnu}.  
The increase of proto-neutron star mass is determined by the profile of progenitor star and the end point for the black hole formation depends on the EOS as in the above case of 40M$_{\odot}$ star and other progenitor examples \citep{sum08}.  
Accordingly neutrino burst is shorter for TM1e than TM1 as shown in Fig. \ref{fig:50MMpnsEnu}.  
Increasing features of average energies are common for all cases, reflecting the continuing accretion and rising temperature.  

\newpage

\subsection{Proto-neutron star cooling} \label{sec:outpnsc}

\begin{figure}[ht!]
\plotone{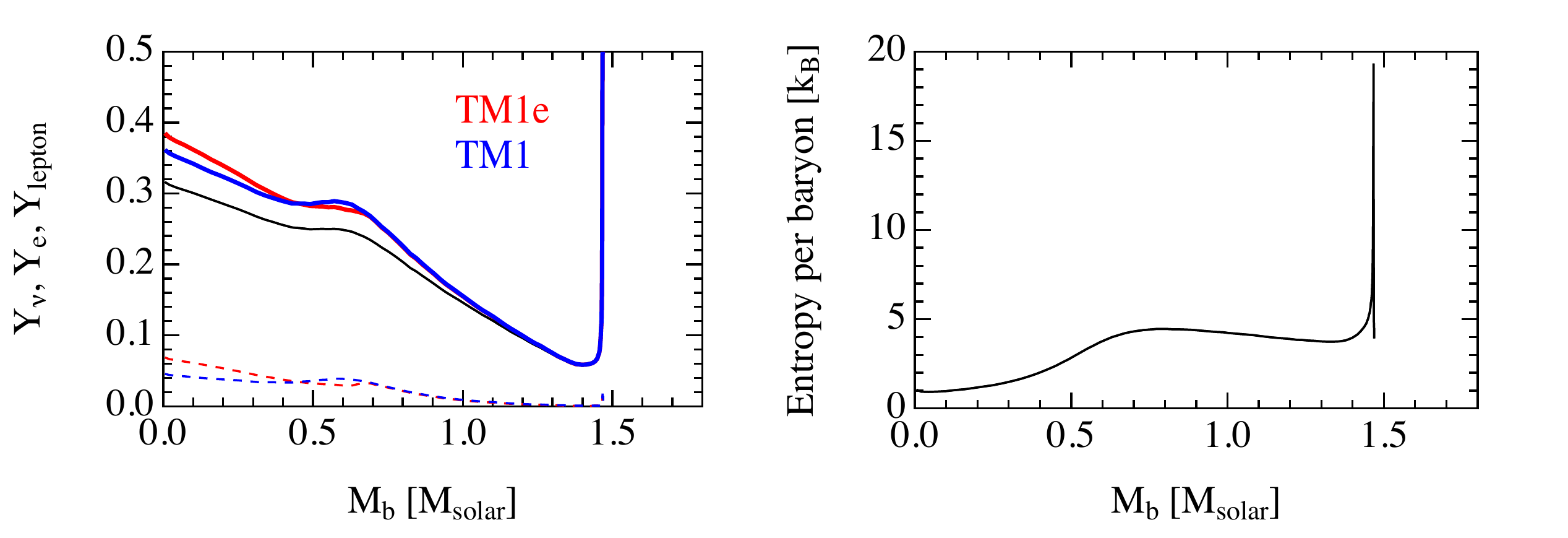}
\caption{Initial profiles of lepton fractions (left) and entropy per baryon (right) in the proto-neutron star cooling are shown as a function of baryon mass coordinate.  The lepton, electron and net neutrino fractions for TM1 (blue) and TM1e (red) models are shown by thick solid, solid and dashed lines, respectively, in the left panel. The distributions of electron fraction and entropy per baryon (black) are used to construct the initial models.   \label{fig:PNSC0sYeS}}
\end{figure}

We show in Fig. \ref{fig:PNSC0sYeS} the initial profiles of lepton fractions and entropy per baryon in the numerical simulations of proto-neutron star cooling.  
The distributions of electron fraction and entropy per baryon are used to construct the initial models for the two sets of EOS, TM1 and TM1e.  
Since we fix the electron fraction as a common value, the lepton fraction for TM1e is slightly larger than that for TM1 having a larger neutrino fraction for TM1e.  
This is due to differences from the symmetry energy since the neutrino fractions are determined under the chemical equilibrium, $\mu_{\nu}=\mu_{e}+\mu_{p}-\mu_{n}$.  
For the case with a small symmetry energy for TM1e at high densities, the difference of chemical potentials between neutrons and protons, $\mu_{n}-\mu_{p}$ becomes small and the resulting $\mu_{\nu}$ becomes large assuming the same electron chemical potential.  

\begin{figure}[ht!]
\plotone{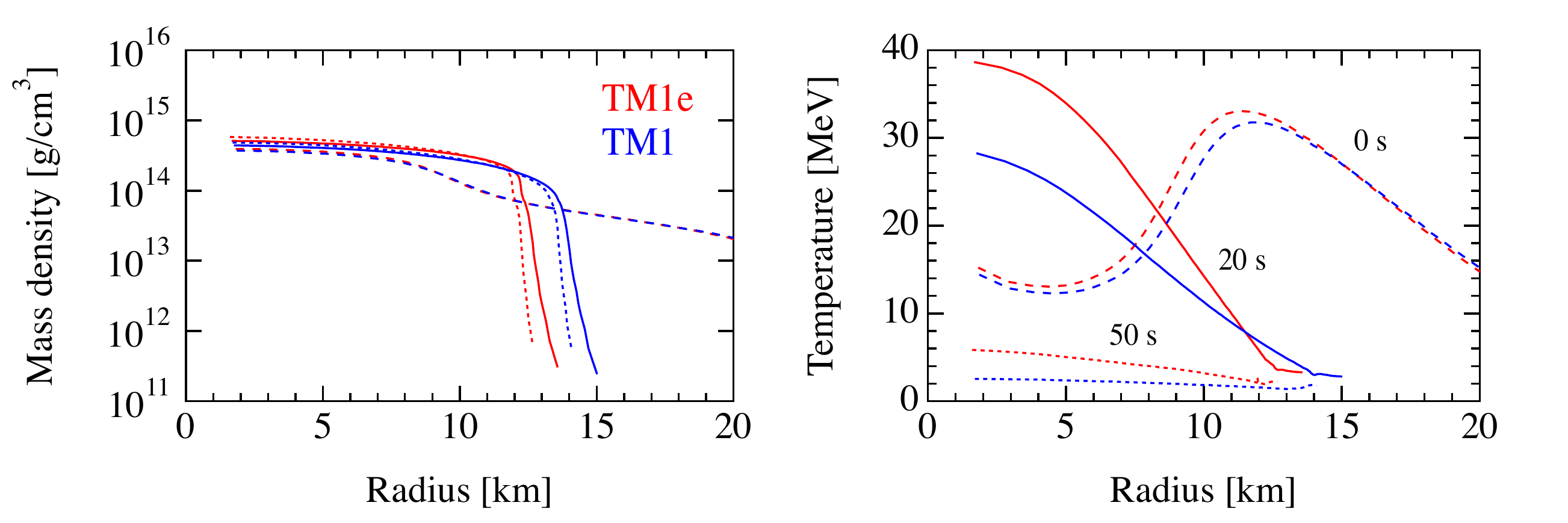}
\caption{Time evolution of profiles of density (left) and temperature (right) in the proto-neutron star cooling is shown as a function of radius.  
Snapshots at 0, 20, 50 s are shown by dashed, solid, dotted lines, respectively, for TM1 (blue) and TM1e (red) models.  \label{fig:PNSCRhoT}}
\end{figure}

Initial profiles constructed by fixing the electron fraction and entropy per baryon are shown in Fig. \ref{fig:PNSCRhoT} for the two models.  
Density profiles are closely similar with a central density of $3.7\times10^{14}$ and $3.9\times10^{14}$ g/cm$^{3}$ for TM1 and TM1e models, respectively.  
The temperature inside 10 km for TM1e is slightly higher than that for TM1.  
The radius of initial proto-neutron star is $\sim$34 km at 10$^{11}$ g/cm$^{3}$ for both cases.  
Therefore, the numerical simulations for the two EOSs start from the very similar profiles.  

\begin{figure}[ht!]
\plotone{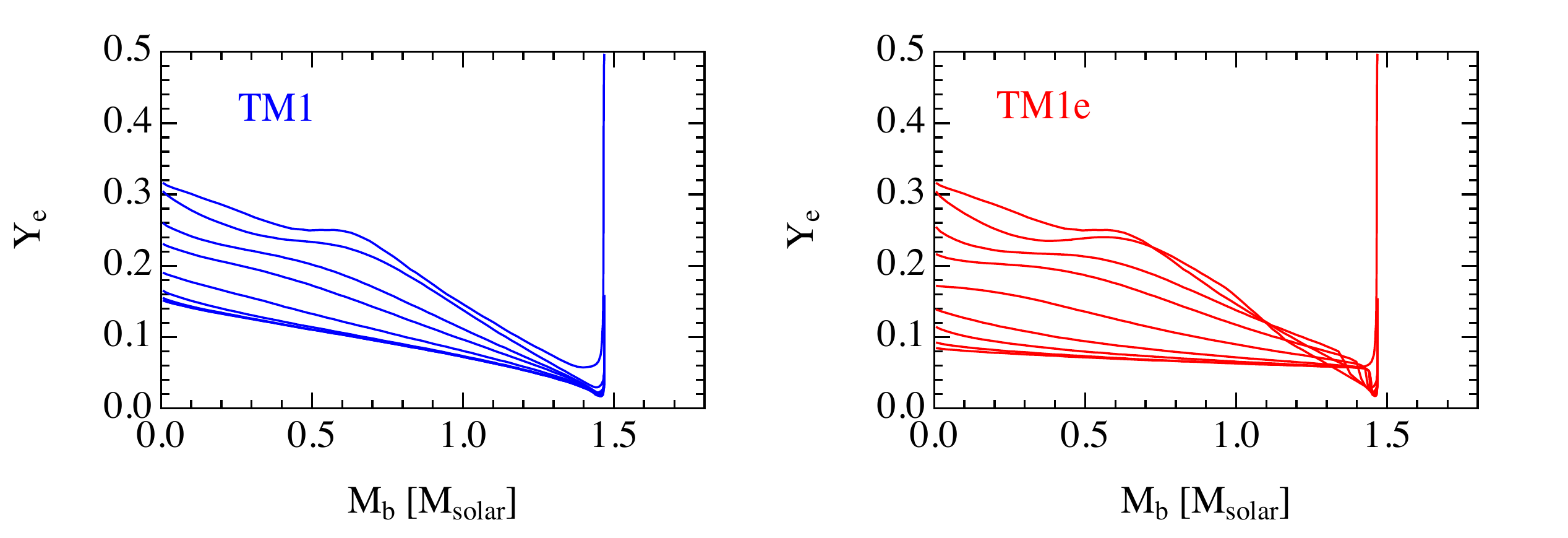}
\plotone{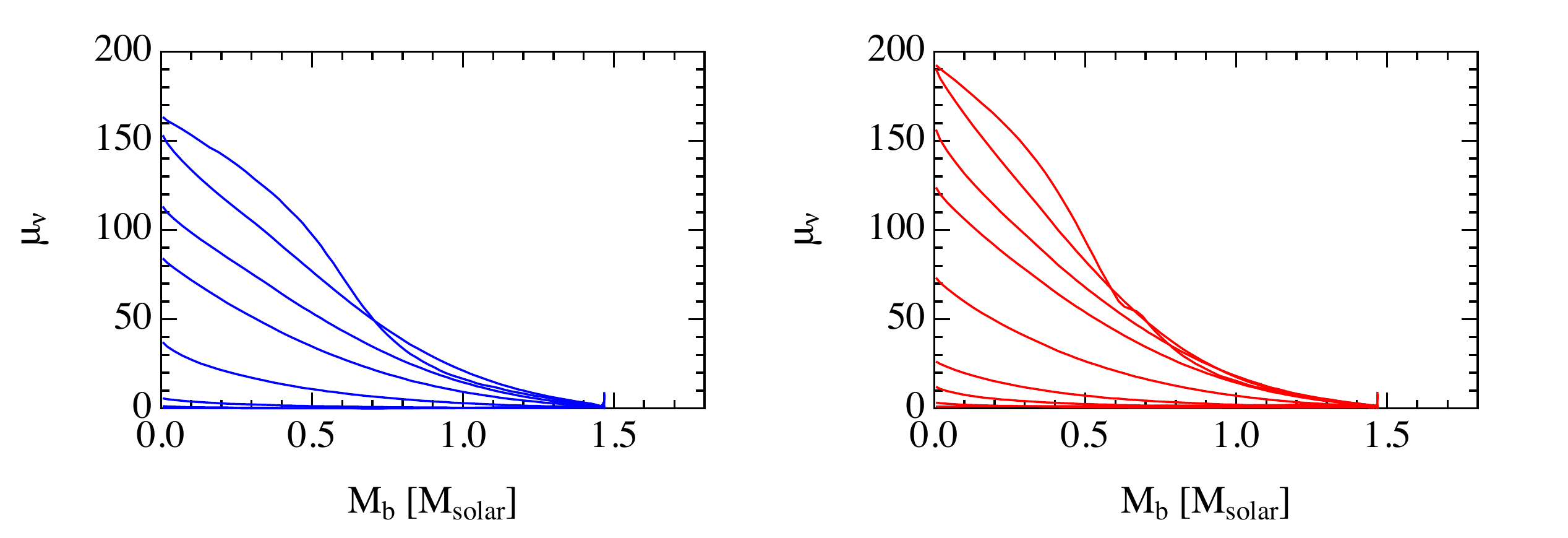}
\caption{Time evolution of profiles of electron fraction (top) and neutrino chemical potential (bottom) is shown as a function of baryon mass coordinate.  
Snapshots at 0, 1, 3, 5, 10, 20, 30, 40, 50 s are shown for TM1 (left) and TM1e (right) models.  \label{fig:PNSCYemunu}}
\end{figure}

\begin{figure}[ht!]
\plotone{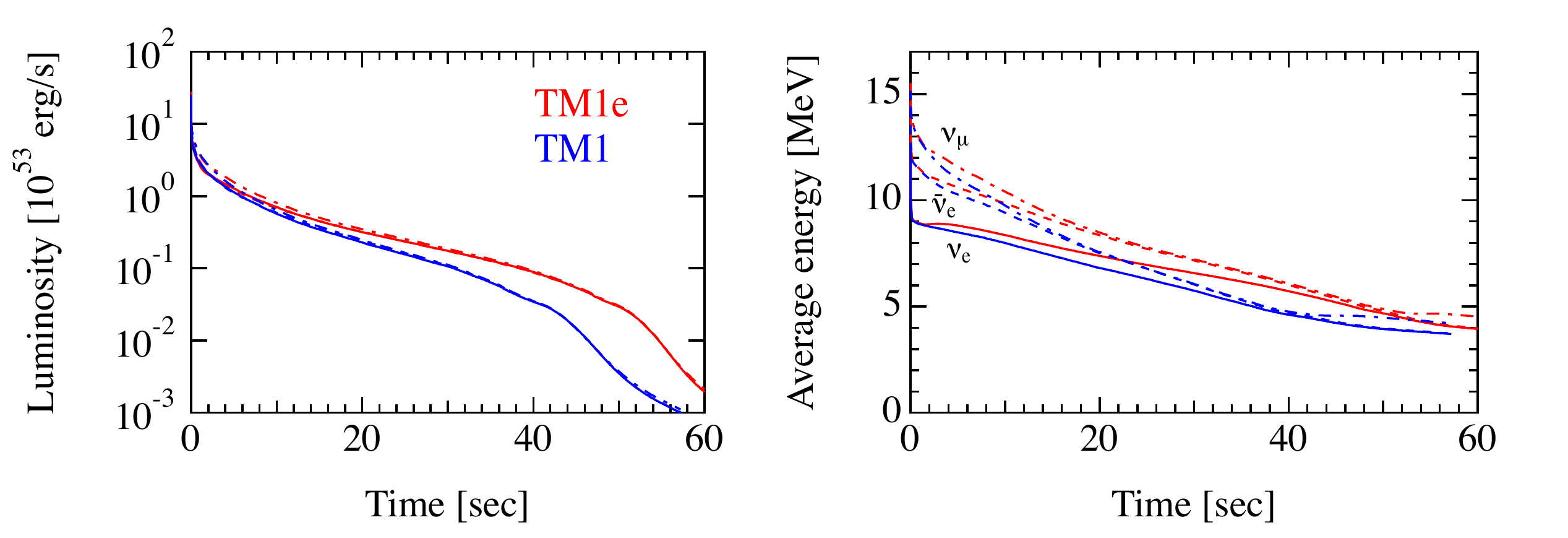}
\caption{Evolution of luminosities and average energies of neutrinos emitted from proto-neutron star as a function of time for TM1 (blue) and TM1e (red) models.  Quantities for the three neutrino species, $\nu_e$, $\bar\nu_e$, $\nu_{\mu}$ are shown by solid, dashed and dash-dotted lines, respectively.  \label{fig:PNSClnuenu}}
\end{figure}

The influence of the EOS appears in a clear manner through the deleptonization in the proto-neutron star cooling.  
As the electron fraction becomes small by neutrino emission (See also Fig. \ref{fig:PNSCYemunu}), effects of different symmetry energy emerge in neutron-rich dense matter.  
The proto-neutron star for TM1e becomes more compact and denser at 20 s having a smaller radius than that for TM1 as seen in the left panel of Fig. \ref{fig:PNSCRhoT}.  
The temperature for TM1e is distinctly higher than that for TM1 at 20 s accordingly.  
The reduction of temperature in the proto-neutron star for TM1e model is large with high luminosities (See Fig. \ref{fig:PNSClnuenu}) as compared with TM1 model.  
Reflecting the different properties of cold neutron stars seen in \S \ref{sec:nstar}, the late phase of proto-neutron stars ends up with different density profiles having the central density at 50 s of $4.9\times10^{14}$ and $5.8\times10^{14}$ g/cm$^{3}$ for TM1 and TM1e models, respectively.  
%

The degree of deleptonization is more drastic in TM1e than in TM1 as can be seen from the comparison of evolution for 50 s in Fig. \ref{fig:PNSCYemunu}.  
The electron fraction decreases from the same initial value and settles down to a smaller value, $Y_e=0.085$, at the center for TM1e than the value, $Y_e=0.15$ for TM1.  
These values nearly correspond to the values from the neutrino-less beta equilibrium for cold neutron stars.  
The drastic deleptonization in TM1e model is apparent having a quick reduction of the neutrino chemical potential as demonstrated in the bottom panels of Fig. \ref{fig:PNSCYemunu}.  
Although the neutrino chemical potential, which is calculated by the proton, neutron and electron chemical potentials assuming the beta equilibrium, for TM1e is larger than that for TM1 at the beginning, it reaches about zero, corresponding to the neutrino-less beta equilibrium, at about 50 s.  
This effect is brought by the difference in the symmetry energies (i.e. difference of chemical potentials between neutron and proton mentioned above) at high densities.  
A large initial value of the neutrino chemical potential at the center leads also to a large gradient of chemical potential, thus enhancing the diffusion for the net fluxes of  electron-type neutrinos for a small symmetry energy.  
The current result of influence is well in accord with the case of linear density dependence of the symmetry energy and for a rapid evolution with long duration of neutrino emission \citep{sum95b}.  

These effects seen in the proto-neutron star evolution reflect the different behavior of symmetry energy at wide densities through the choice of $L$, which is correlated with the value of $E_{sym}$ in the RMF theory.  In the case of TM1e, the small values of $L$ and $E_{sym}$ at the saturation density lead to small symmetry energy at high densities.  This brings the softness of EOS, the large neutrino chemical potential and the small proton fraction in cold neutron stars.  They in turn lead to high density and temperature as well as a large gradient and change of neutrino chemical potential in the proto-neutron star.  It is interesting that both TM1 and TM1e have the same symmetry energy at 0.11 fm$^{-3}$, which is slightly lower than the saturation density for the description of finite nuclei, but they show different consequences. Future experiments on the neutron skin of heavy nuclei to precisely determine the value of $L$ are helpful in this regard.   Incidentally, the symmetry energy at twice the saturation density, which is important to determine the neutrino cooling time scale \citep{nak19}, is 74.7 MeV and 46.0 MeV for TM1 and TM1e, respectively.  %

The differences of thermal evolution seen above appear through different neutrino emission as shown in Fig. \ref{fig:PNSClnuenu}.  
The luminosity for TM1e model becomes higher than that for TM1 due to high temperature during the evolution over 50 s.  
The decay of luminosity is slow for TM1e since the density of proto-neutron stars is high and the diffusion time scale becomes long through short mean free paths.  
In TM1e model, 
the change of proto-neutron star through cooling is drastic with high luminosities by emitting a larger amount of thermal energy and lepton number.  
The average energies of neutrinos for TM1e are higher due to higher temperature and neutrino chemical potential of dense matter than those for TM1.  
High average energies of neutrinos may also contribute to an increase of the opacity for diffusion.  

The future observation of a supernova neutrino burst may be able to examine this difference of neutrino emissions.  In the case of detection of a Galactic supernova at the Super-Kamiokande detector, the total event number is evaluated to be $\sim10^{4}$ and a long duration of neutrino burst over 50 sec can be observed \citep{suw19}.  The detection ability in principle enables us to distinguish the different decreasing features of neutrino luminosity between TM1 and TM1e.  The difference in average energies of 1--2 MeV may be marginally examined at Super-Kamiokande and will be clarified at Hyper-Kamiokande \citep{hypk18}.  It should be stressed, however, that the neutrino emission depends also on the properties of compact object and the other parameters of EOS.  The time profile of neutrino burst, for example, depends on the mass of proto-neutron star through the total binding energy.  The luminosities and average energies reflect the density and temperature of dense matter, which are affected by the stiffness and other parameters of EOS such as effective mass besides the symmetry energy \citep{nak19}.  The current study reveals the size and direction of differences due to the symmetry energy.  Further studies on the difference in neutrino detection are definitely necessary by covering a variety of neutron star mass.  




\section{Discussion} \label{sec:discussion}

Although the effect of symmetry energy on the post-bounce evolution in core-collapse supernovae is found to be minor due to a less neutron-rich environment, further studies are necessary to judge the consequence of supernova explosions.  
First of all, the current study is limited to the effect of uniform matter at high densities.  
The effects of low density part of the non-uniform matter with the small $L$ value will be examined in numerical simulations by future studies with the construction of the modified Shen EOS table, which is currently underway.  
The small symmetry energy may alter the fraction of free protons and nuclei under the nuclear statistical equilibrium by shifting the chemical potentials.  
Such changes may lead to different reaction rates of neutrino emission and absorption.  
Electron capture rates during the gravitational collapse have influence on the neutrino trapping and the properties of central core at the bounce as an initial configuration of shock propagation \citep{bru89a,swe94}.  
A different composition of nucleons and nuclei including light ones at low densities may affect the neutrino emission and absorption for neutrino heating \citep{fur13b,fis16,nag19a}.  
Secondly, the current study is done under the spherical symmetry.  
Although the effects of EOS in multi-dimensional dynamics of core-collapse supernovae have been studied \citep{mar09b,jan12a,suw13,cou13,fis14,pan18,nag18}, the adopted EOS tables are limited and not systematic having various bulk properties.  
Recent systematic studies suggest the effective mass provides the largest uncertainty in the neutrino signal 
and it is influential for explosion \citep{yas18,sch19b}.  
Note that the effective mass for TM1e is the same as that for TM1 since there is no isovector-scalar interaction involved in the extension.  
It would be interesting to take out 
the sole effect of the symmetry energy in multi-dimensional supernovae by adopting the new EOS with the small $L$ value since the properties of symmetric matter are preserved as in the Shen EOS.  

The minor effect in supernova cores found in the current study suggests the importance of careful study on the bulk parameters of nuclear matter in supernovae (See also \citet{sch19b}).  
Although the change of symmetry energy is large having different behavior of neutron matter, 
properties of less neutron-rich dense matter in supernova cores are similar between TM1 and TM1e EOSs.  
In this sense, it is important to explore effects of nuclear matter at $Y_p\sim0.3$ just above the nuclear matter density. 
Although the EOS of neutron matter is getting more constrained by nuclear experiments, neutron star observations and gravitational wave detection, further efforts on nuclear matter close to the symmetric situation remain essential through theoretical and experimental approaches in nuclear physics.   

As we have seen the effect in the late phase of proto-neutron stars, the symmetry energy is essential to determine the final form of a neutron star or black hole.  
From the similar properties of central core around the bounce with different symmetry energies, the outcome of long term evolution over 1 s may turn out to be different through configuration of cold neutron stars and the threshold to the black hole formation. 
The resulting emission of neutrinos at the late stage is affected since the final form of central object is largely determined by the symmetry energy (See also \citet{rob12} on the effects through convection).  
The total energy of supernova neutrinos is accordingly affected by the symmetry energy through the properties of a cold neutron star.  
The information on symmetry energy may be derived from the future detection of a supernova neutrino burst.  

Since the supernova neutrino database \citep{nak13a} is constructed by using the Shen EOS, it would be preferable to handle the data of neutrino emission with an expected modifications for predictions of neutrino bursts and diffuse supernova neutrino background.  
Since the difference in symmetry energy has a minor influence around the core bounce and provides  distinguishable differences in proto-neutron stars, we expect that the data of neutrino emission for the early phase remain similar and those for the late phase may have slight enhancement in energy and flux with TM1e EOS.  
Systematic studies on the EOS dependence in the cooling of proto-neutron stars have been done \citep{nak18,nak19} and will be reported elsewhere.  

\section{Summary} \label{sec:summary}

We study the influence of density dependence of symmetry energy at high densities in the numerical simulations of gravitational collapse of massive stars and cooling of proto-neutron stars.  
We utilize the revised table of equation of state (EOS) by the relativistic mean field (RMF) theory based on the Shen EOS table, which has been routinely used for astrophysical simulations.  
In order to take out the symmetry energy effect by separating out other modifications, we prepare a modified EOS table with a small symmetry energy and apply it to supernova simulations.  

The RMF theory is extended by introducing the $\omega-\rho$ coupling term 
to describe the density-dependent symmetry energy.  
This extension modifies the dependence of the symmetry energy in the RMF with TM1 interaction for the Shen EOS table, which is claimed to be too strong in the current knowledge of constraints, under the exactly same behavior of symmetric nuclear matter.  
Based on the systematic studies on the value of $L$, 
we adopt the case of $L=40$ MeV (denoted as TM1e), which is in accord with the current experimental and observational data, among them.  

We utilize the same RMF theory to construct the EOS table of uniform matter to cover the wide conditions at finite temperature.  
We adopt the TM1e EOS beyond the density of 10$^{14}$ g/cm$^3$ by switching from the TM1 EOS in the Shen EOS.  
By keeping the low density part of non-uniform matter as in the Shen EOS, we explore the influence of different behavior of symmetry energy at high densities.  
We perform new simulations with the TM1e EOS table keeping other settings and compare with the previous results obtained by using the Shen EOS.  

We found similar evolution around the core bounce and a shorter evolution toward the black hole formation in numerical simulations of neutrino-radiation hydrodynamics for the gravitational collapse of massive stars.   
In the case of 11.2M$_{\odot}$ and 15M$_{\odot}$ stars, the profiles of central core in the TM1 and TM1e cases are closely similar around the core bounce because the proton fraction is rather large (less neutron-rich) and the density is not so high up to 200 ms after bounce.  
Associated properties of neutrino emission are essentially the same in the early post-bounce phase.  
In the case of 40M$_{\odot}$ and 50M$_{\odot}$ stars, where the mass accretion is intense and non-explosion is inevitable, 
the difference appears distinctively at $\sim$1 s having the higher density and temperature in increasing mass of proto-neutron stars.  
The duration of the increasingly massive proto-neutron star for TM1e is shorter than that for TM1 reflecting the different stiffness to determine the maximum mass.  
The neutrino burst continues up to the black hole formation and, therefore, is shorter for TM1e than for TM1.  

We found gradual growth of EOS difference in  the cooling of proto-neutron stars by numerical simulations of the quasi-static thermal evolution with neutrino emission.  
The proto-neutron star in TM1e becomes more compact and hot as compared with the case of TM1 during the evolution over 20 s.  
As the thermal evolution proceeds through the deleptonization by neutrino emission, the dense matter becomes neutron-rich and the EOS influence gradually appears reflecting the different symmetry energy.  
The evolutionary change of proto-neutron stars is more drastic in TM1e through larger neutrino emissions than in TM1.  
The neutrino luminosities and average energies in TM1e are higher than those in TM1, leading to the larger total energy and number of neutrino emission with a large change of configuration in TM1e.  
The decay of neutrino luminosities is slow, in contrast, due to the high density in TM1e through the different time scale of diffusion.  

The full table of the modified Shen EOS with TM1e including the low density part is under construction.  
It would be interesting to study the influence of symmetry energy in non-uniform matter at low densities on the core-collapse supernovae and proto-neutron stars.  
Symmetry energy effects in multi-dimensional simulations of supernovae remain to be explored as applications of the TM1e EOS.  
The current study will be the basis to discuss the influence further.  
It is also fascinating to perform other astrophysical simulations including neutron star mergers having the TM1e EOS, which is compatible with cold neutron star profiles.  
\acknowledgments

K. S. would like to thank H. Togashi, S. Furusawa, Y. Sekiguchi, Y. Suwa, K. Oyamatsu and H. Toki for fruitful discussions and suggestions on EOS tables and their applications.  
This work is supported by
Grant-in-Aid for Scientific Research
(15K05093, 19K03837), 
Grant-in-Aid for Scientific Research on Innovative areas 
"Gravitational wave physics and astronomy:Genesis"
(17H06357, 17H06365)
and 
Grant-in-Aid for Scientific Research on Innovative areas 
"Unraveling the History of the Universe and Matter Evolution with Underground Physics" 
(19H05802, 19H05811) 
from the Ministry of Education, Culture, Sports, Science and Technology (MEXT), Japan.  
This work is also supported in part by the National Natural Science Foundation of China (Grants No. 11675083 and No. 11775119).
For providing high performance computing resources, 
Computing Research Center of KEK, 
JLDG on SINET4 of NII, 
Research Center for Nuclear Physics, Osaka University, 
Yukawa Institute of Theoretical Physics, Kyoto University, 
Nagoya University, 
Hokkaido University
and 
University of Tokyo are acknowledged. 
This work was partly supported by 
research programs at K-computer of the RIKEN AICS, 
HPCI Strategic Program of Japanese MEXT, 
“Priority Issue on Post-K computer” (Elucidation of the Fundamental Laws and Evolution of the Universe)
and 
Joint Institute for Computational Fundamental Sciences (JICFus).

\end{document}